\documentclass[12pt]{article}
\usepackage{multirow}
\usepackage{longtable}
\usepackage{booktabs}
\usepackage{geometry}
\usepackage{graphicx}
\usepackage{setspace}      
\usepackage{amssymb}
\usepackage{comment}
\usepackage{amsmath,bm}
\usepackage{mathrsfs}
\usepackage{amsthm}
\usepackage{amsfonts}
\usepackage{amssymb}
\usepackage{amsmath}
\usepackage{subfigure}
\usepackage{floatrow}
\usepackage{epsfig,amssymb,latexsym,verbatim}
\usepackage{graphicx}
\usepackage{epstopdf}
\usepackage{multirow}
\usepackage{graphics}
\usepackage{amssymb}
\usepackage{amsthm}
\usepackage{multirow}
\usepackage{booktabs}
\usepackage[english]{babel}
\usepackage[figuresright]{rotating}
\newtheorem{theorem}{Theorem}

\newtheorem{corollary}{COROLLARY}

\usepackage{epsfig,amsmath,float}
\usepackage{graphics}
\usepackage[dvipsnames]{xcolor}
\usepackage{epsfig}
\usepackage{subfigure}
\usepackage{color}
\usepackage{amssymb}
\usepackage{setspace}
\usepackage{hyperref}
\usepackage{float}

\usepackage{mathrsfs}

\def\E{\mathbb{E}}

\def\G{\mathcal{G}}
\def\J{\mathcal{J}}

\def\Var{{\rm{Var}}}
\def\Cov{{\rm{Cov}}}
\def\P{\mathbb{P}}
\def\O{\mathcal{O}}
\def\o{{\scriptstyle{\mathcal{O}}}}

\hypersetup{ colorlinks = true, linkcolor = blue, citecolor = blue}  
\usepackage[square,sort,comma,numbers]{natbib}
\renewcommand{\citep}[1]{\citeauthor{#1}, \citeyear{#1}}  

\numberwithin{equation}{section}

\title{From Sparse to Dense Functional Data: Phase Transitions from a Simultaneous Inference Perspective}
\author{Leheng Cai$^1$ and Qirui Hu$^1$\thanks{The corresponding author. E-mail address: hqr20@mails.tsinghua.edu.cn} 
	\\~\\
	{\small \it $^{1}$  Center for Statistical Science and Department of Industrial Engineering,}
	\\{\small \it Tsinghua University, Beijing, China}
}
\date{}
\begin{document}
	\maketitle
	\begin{abstract}
		We aim to develop  simultaneous inference tools for the mean function of functional data from sparse to dense. First, we derive a unified Gaussian approximation to construct simultaneous confidence bands of mean functions based on the B-spline estimator. Then, we investigate the conditions of phase transitions by decomposing the asymptotic variance of the approximated Gaussian process.  As an extension, we also consider the orthogonal series estimator and show the corresponding conditions of phase transitions. Extensive simulation results strongly corroborate the  theoretical results, and also illustrate the variation of the asymptotic distribution via the asymptotic variance decomposition we obtain. The developed method is further applied to body fat data and traffic data.   
	\end{abstract}
	\noindent%
	\vfill
	
	\section{Introduction}

	\par Over the past two decades, functional data analysis has evolved significantly, gaining prominence in diverse fields such as environmental science, industry, and neuroscience. Key resources for foundational concepts in this area include \cite{bosq2000linear}, \cite{Ferraty2006NonparametricFD}, and \cite{hsing2015theoretical}.
	
	\par Central to functional data analysis is the nonparametric estimation of mean and covariance functions from discretely sampled curves with noise. These estimations are crucial not just as standalone analysis, but also as key components in dimension reduction and further modeling of functional data. This includes applications in functional principal component analysis; see \cite{yao2005functional}, \cite{Li2010UniformCR},  and \cite{CAI2024107900}, functional linear regression; see \cite{Yao2005FunctionalLR}, and change point analysis; see \cite{Berkes2009DetectingCI}. 
	
	\par In standard functional data scenarios, we typically encounter $n$ random curves, each representing a subject, with measurement errors at $N_i$ random time points for the $i$-th subject. The sampling frequency, denoted as $N_i$, plays a pivotal role in the selection of appropriate estimation procedures. Research in this area generally falls into two primary categories: sparse and dense functional data. For sparse functional data, marked by boundedness of $N_i$, one often   pools data across subjects to gain efficiency, as discussed in \cite{yao2005functional} and \cite{Li2010UniformCR}, since pre-smoothing is not feasible.   On the other hand, for dense functional data, characterized by $N_i$ significantly surpassing a certain threshold relative to $n$, one typically adopts nonparametric smoothing for each subject's data to reduce noise and reconstruct individual curves, as seen in \cite{wang2020simultaneous} and \cite{2022lijie}.
	
	Furthermore, \cite{ZW16} performed an in-depth analysis of phase transitions and meticulously investigated convergence rates from sparse to dense functional data, introducing additional subcategories like “semi-dense” and “ultra-dense”. These are based on whether achieving the $\sqrt{n}$ rate and ignoring the asymptotic bias. Also, \cite{sharghi2021mean} investigated the estimation of mean derivatives in longitudinal and functional data from sparse to dense. Studies such as \cite{tonycai2011} and \cite{berger2023dense} have explored the optimal rates for estimating the mean function in  the $\mathcal L^2$-norm and sup-norm, respectively. More recent work by \cite{Zhang2021UnifiedPC} delved into unified principal component analysis for sparse and dense functional data, especially under spatial dependency conditions. Lastly, \cite{guo2023sparse} achieved significant breakthroughs in non-asymptotic mean and covariance estimation within high-dimensional functional time series. Their contribution is particularly noteworthy for establishing  rates for both $\mathcal L^2$ convergence and uniform convergence, offering powerful tools for convergence analysis when the dimensionality of functional data grows exponentially compared to the sample size.

	\par However, existing asymptotic results mainly address consistency issues, including $\mathcal L^2$ and uniform convergence, or local asymptotic normality. Simultaneous confidence inference is a vital aspect of functional data analysis. It focuses on the distribution of uniform deviations rather than pointwise asymptotic distributions for constructing  confidence intervals. Nonparametric simultaneous confidence bands (SCBs) are essential for simultaneous inference on functions. In dense data contexts, SCBs have been constructed for mean functions; see \cite{cao2012simultaneous}, \cite{2022lijie}, \cite{Hu2022}, \cite{HuStatisticalIF}, \cite{huang2022inference}, covariance functions; see  \cite{cao2016oracle}, \cite{wang2020simultaneous}, \cite{HL23}, and functional principal components; see \cite{CAI2024107900}, \cite{CH23}. \cite{caiHuFPCscore} introduced a multi-step smoothing method for SCBs of distribution functions of FPC scores. In contrast, the sparse data setting poses challenges for simultaneous inference due to issues like non-tightness of estimators  and  lack of weak convergence. To the best of our knowledge, \cite{zheng2014smooth} is the only  work providing the SCBs of mean function on interval $[h,1-h]$ with bandwidth $h$ under sparse sampling, while their method assume the distribution of FPC scores are i.i.d. and follow Gaussian distribution, and the local linear estimator will suffer the boundary problem leading to the SCBs cannot hold on the entire interval $[0,1]$.
	
	\par This paper introduces a novel methodology for simultaneous inference of the mean function under arbitrary sampling strategies. We propose a B-spline estimator to overcome the shortcomings of kernel smoothing, such as boundary problems and low computational efficiency. Our approach includes a unified Gaussian approximation, facilitating constructing SCBs  for mean functions of functional data from sparse to dense. By decomposing asymptotic variance, we observe phase transitions in the approximated Gaussian process. As the number of observation points increases with sample size, we identify three stages of transition, namely ``sparse'',       ``semi-dense'', and   ``dense'', similar to \cite{ZW16}'s findings. For sparse data, the covariance function of the Gaussian process involves the diagonal of the covariance function of trajectories and the variance function of observation errors. 
	The estimator itself is 
	lack of asymptotic equicontinuity, resulting in  no limiting distribution in $C[0,1]$. 
	In dense  scenarios, the effect of measurement errors becomes negligible.
	There exists an intermediate regime between the
	sparse and dense cases, referred as ``semi-dense'', dominated by the contribution of both trajectories and   errors, which coincides with results in \cite{berger2023dense}. 
	Our findings align with those of \cite{ZW16} under specific assumptions, offering more general results relying on the moment of trajectories and the smoothness of the mean function. Furthermore, we establish a sufficient condition for ``ultra-dense'' sampling, achieving “oracle” efficiency as described in \cite{cao2012simultaneous}, where the estimator is asymptotically equivalent to all the trajectories that can be completely recorded   without errors. Additionally, we extend our methodology to general orthogonal series estimators, presenting novel results for simultaneous inference of mean functions from sparse to dense.
	
	\par The remainder of the paper is structured as follows: Section 2 details the B-spline estimator and its asymptotic properties. Section 3 expands on this framework for orthogonal series estimators. Sections 4 and 5 present implementation details and simulation results, respectively. Application for two real world datasets is illustrated in Section 6, and all proofs of main results are included in supplemental material.
	

	\section{Main results}
	\subsection{Notations}
	\par Before we describe our methodology, some notations are introduced first.  For any vector $ \bm{a} = \left(a_1,\ldots,a_s  \right)^\top \in \mathbb{R}^s $, take $ \left\Vert \bm{a}\right\Vert_r = \left( \left| a_1\right|^r+ \ldots+\left| a_s\right|^r \right)^{1/r}$ for $1 \le r < \infty$, and $ \left\Vert \bm{a} \right\Vert_\infty=\max_{1\leq j\leq s}|a_{j}|$. For any $n\times n$ matrix $\mathbf A=(a_{ij})_{i,j=1}^{n}$, denote its $L_r$-norm as $\left\|\mathbf A\right\|_{r}=\sup_{\left\|x\right\|_r=1}\left\|\mathbf Ax\right\|_{r}$ for $1\leq r \le \infty$, and particularly $\left\|\mathbf A\right\|_{op}=\left\|\mathbf A\right\|_{2}$ and $\left\|\mathbf A\right\|_{\max}=\max_{1\leq i,j\leq n}|a_{ij}|$. For any function $f(\cdot)$ defined on domain $\mathcal D$, denote its sup-norm as $\left\|f\right\|_{\infty}=\sup_{x\in\mathcal D}\left|f(x)\right|$, and denote its $\mathcal{L}^2$-norm as $\left\|f\right\|_{\mathcal{L}^2}=\left\{\int_{\mathcal D} f^2(x)dx\right\}^{1/2}$. 
	For a real number $\nu\in(0,1]$ and integer $q\in\mathbb{N}$, write $\mathcal{H}_{q,\nu}(\mathcal D)$ as the space of $(q,\nu)$-H\"{o}lder continuous functions on $\mathcal D$, that is,
	\begin{align*}
		\mathcal{H}_{q,\nu}(\mathcal D)=\left\{h:\mathcal D\to \mathbb{R},~\left\|h\right\|_{q,\nu}=\sup_{t,s\in\mathcal D,t\neq s,\left\|\bm\alpha\right\|_1=q}\frac{\left|\partial^{\bm \alpha} h(t)-\partial^{\bm \alpha} h(s)\right|}{\left\|t-s\right\|_2^{\nu}}<\infty \right\}.
	\end{align*}
	For real numbers $a$ and $b$, $a\wedge b$, $a\vee b$, $\lfloor a\rfloor$ and  denote $\min\{a,b\}$, $\max\{a,b\}$ and  largest integer that is smaller or equal to $a$, respectively.
	For two sequences $\{a_n\}$ and $\{b_n\}$, we write $a_n\asymp b_n$, $a_n\lesssim b_n$, $a_n\gtrsim b_n$, $a_n\ll b_n$ and $a_n\gg b_n$ to mean
	that there exist $c$, $C>0$ such that $0<c\leq |a_n/b_n|\leq C<\infty$,  $a_n\leq  Cb_n$,  $b_n\leq  Ca_n$, $a_n=\o(b_n)$, and $b_n=\o(a_n)$,
	respectively. For random variables $X$ and $Y$, we say $X\overset{d}{=}Y$ if they have the same distribution. 
		
	\subsection{B-spline estimator}
	\par For illustration, and without loss of generality, we assume that the domain of the functional data $\eta(x)$ is the unit interval $[0,1]$, which is a square-integrable continuous
	stochastic process: specifically, $\eta \left(x\right) \in \mathcal{L}^2
	[0,1]$ almost surely, satisfying $\mathbb{E}\int_0^1\eta^2
	(x)dx<\infty $, with mean function $m(x)$ and covariance function $G(x,x^{\prime})$. According to \cite{hsing2015theoretical}, there exist eigenvalues $\lambda_{1}\geq \lambda_{2}\geq \cdots \geq 0$ with $\sum_{k=1}^{\infty }\lambda_{k}<\infty $, and
	corresponding eigenfunctions $\left\{ \psi_{k}\right\} _{k=1}^{\infty }$ of
	$G\left( x,x^{\prime}\right) $  being an orthonormal basis of $
	\mathcal{L}^2\left[ 0,1\right] $,  such that $G\left( x,x^{\prime }\right)
	=\sum_{k=1}^{\infty }\lambda _{k}\psi_{k}(x)\psi _{k}\left( x^{\prime
	}\right)$, $\int G\left( x,x^{\prime }\right) \psi_{k}\left(x^{\prime
	}\right) dx^{\prime }=\lambda_{k}\psi_{k}(x)$. The demeaned process $\chi_i(x)=\eta_i(x)-m(x)$, then allows the well-known Karhunen-Lo\`{e}ve representation $\chi_i (x)=\sum_{k=1}^{\infty }\xi_{ik}\phi_{k}(x)$, in which the random coefficients $\left\{ \xi_{ik}\right\}_{k=1}^{\infty }$, called functional principal component (FPC) scores, are
	uncorrelated with mean $0$  and variance $1$. The rescaled eigenfunctions $
	\phi_{k}(x)$, called FPCs, satisfy that $\phi _{k}(x)=\sqrt{\lambda_{k}}\psi
	_{k}(x)$, for $k\in\mathbb Z_+$.
	
	\par  Consider the following model:
	\begin{align}\label{model1}
		Y_{ij} &=\eta_i(X_{ij})+\sigma(X_{ij})\epsilon_{ij}, \quad 1\leq i\leq n, \,\,1\leq j\leq N_i.
	\end{align}
	where   covariates $X_{ij}$'s are i.i.d. with density $f(\cdot)$,   $\epsilon_{ij}$'s are independent random variables with $\E\left(\epsilon_{ij}\right)=0$ and $\Var\left(\epsilon_{ij}\right)=1$, and $\sigma(\cdot)$ is the variance function of measurement errors. 
	According to the Karhunen-Lo\`{e}ve  representation, 
	one could further rewrite model (\ref{model1}) as \begin{align}\label{model2}
		Y_{ij} =m(X_{ij})+\sum_{k=1}^{\infty}\xi_{ik}\phi_k(X_{ij})+\sigma(X_{ij})\epsilon_{ij}, \,\, 1\leq i\leq n, \,\,1\leq j\leq N_i, 
	\end{align}
	and $\{\xi_{ik}\}_{i=1,k=1}^{n,\infty}$, $\{\epsilon_{ij}\}_{i=1,j=1}^{n,N_i}$ and $\{X_{ij}\}_{i=1,j=1}^{n,N_i}$ are mutually independent. 
	\par We first consider a B-spline approach to estimate and conduct simultaneous inference for the mean function $m(x)$. 
	To describe the spline functions, let $\{t_l\}_{l=0}^{J_n+1}$ be a sequence of equally-spaced points, where $t_l = l / (J_n+1)$ for $0 \leq l \leq J_n+1$, which divide $[0,1]$ into $J_n+1$ equal sub-intervals, denoted as $I_l = [t_l, t_{l+1})$ for $l = 0, \ldots, J_n-1$, and $I_{J_n} = [t_{J_n}, 1]$. Let $\mathcal{S}^{p}_{J_n} = \mathcal{S}^{p}_{J_n}[0,1]$ be the polynomial spline space of order $p$ over $ 
	\left\{I_l\right\}_{l=0}^{J_n}$, consisting of all functions that are $(p-2)$ times continuously differentiable on $[0,1]$ and are polynomials of degree $(p-1)$ within the sub-intervals $I_l$, $l = 0,\ldots,J_n$. With a little abuse of notation, let $\{B_{l,p}(x)\}_{1 \leq l \leq J_n + p}$ be the $p$-th order B-spline basis functions of $\mathcal{S}_{J_n}^{p}$, then $\mathcal{S}_{J_n}^{ p}=\left\{
	\sum_{l =1}^{J_n+p}\lambda _{l}B_{l,p}(x
	): \lambda _{l}\in \mathbb{R} \right\} $. 
	
	The mean function $m(x)$ can be  estimated by solving the following spline regression, defined as:
	\begin{equation}  \label{DEF:mhat}
		\hat{m}(x)=\mathop{\arg\min}
		\limits_{g(\cdot)\in \mathcal{S}_{J_n}^{p}}
		\sum_{i=1}^n\sum_{j=1}^{N_i}  \left\{Y_{ij}-g
		(X_{ij})\right\}^2.
	\end{equation}
	Denote $\bar N=n^{-1}\sum_{i=1}^{n}N_i$, and $p$-th order B-spline basis by $\bm B(x)=\bm B_{J_n,p}(x)=\left( B_{1,p}(x),\ldots, B_{J_n+p,p}(x)\right)^\top$ . The solution to (\ref{DEF:mhat}) is $\hat m(x)=\bm B^\top (x)\hat{\bm\theta}$, in which \begin{align*}
		&\hat{\bm\theta} =\hat{ \mathbf{V}}^{-1}\left\{\frac{1}{n\bar N}\sum_{i=1}^{n}\sum_{j=1}^{N_i} \bm B(X_{ij})Y_{ij}\right\},\quad 
		\hat{\mathbf{V}}= \frac{1}{n\bar N}\sum_{i=1}^{n}\sum_{j=1}^{N_i} \bm B(X_{ij}) \bm B^\top (X_{ij}).
	\end{align*}
	
	\subsection{Strong approximation by Gaussian processes}\label{sec2.1}
	
	\par In this subsection, we investigate asymptotic properties of the B-spline estimator in (\ref{DEF:mhat}). Without loss of generality, we assume that $N_i\geq 2$ for convenience to describe and estimate the covariance structure in the following. Some technical assumptions are introduced for theoretical development. 
	{\color{black}
		\begin{itemize}
			\item[(A1)] Assume that $m(\cdot)\in \mathcal{H}_{q,\nu}[0,1]$ for some integer $q>0$ and some positive real number $\nu\in(0,1]$. In the following, denote by $q^\ast=q+\nu$.
			\item[(A2)] The density function $f(\cdot)$ is bounded above and away from zero, i.e., $c_{f}\leq \inf_{x\in[0,1]} f(x)  \leq \sup_{x\in[0,1]} f(x) \leq C_{f}$ for some positive constants $c_{f}$ and $C_{f}$.
			\item[(A3)] 
			The variance function $\sigma(\cdot)$ of measurement error is uniformly bounded, and  $\E|\epsilon_i|^{r_1}<\infty$ for some integer $r_1\geq 3$.
			\item[(A4)] The covariance function $G(\cdot,\cdot)\in\mathcal{H}_{0,\beta}[0,1]^2$ for some $\beta\in (0,1]$,   $ \inf_{x\in[0,1]}\\G(x,x)\geq c_{G}$ for some positive constant $c_{G}$, and $\sup_{x\in[0,1]}\E|\eta(x)|^{r_2}<\infty$ for some integer $r_2\geq 3$.
			\item[(A5)] Suppose that the spline order $p\geq q^\ast$. Let  $r=\min\{r_1,r_2\}$, When $\bar N\lesssim J_n$, $J_n^{-q^\ast-1/2}n^{1/2}\bar N^{1/2}\log^{1/2}(n)=\o(1),$
			\begin{align*}
				J_n^{3-2/r} n^{-2+2/r}\bar N^{-1}\log^2(n)\sum_{i=1}^{n}\left(\sum_{s=1}^{r}N_i^s J_n^{-s}\right)^{2/r}=\o(1).
			\end{align*}
			When $\bar N\gg J_n$, $J_n^{-q^\ast}n^{1/2}\log^{1/2}(n)=\o(1),$
			\begin{align*}
				J_n^{4-2/r} n^{-2+2/r}\bar N^{-2}\log^2(n)\sum_{i=1}^{n}\left(\sum_{s=1}^{r}N_i^s J_n^{-s}\right)^{2/r}=\o(1).
			\end{align*}
			\item[(A5')] Let $r=r_2$ in (A5).
			
		\end{itemize}
	}

	These assumptions are rather mild.   Assumptions (A1) and (A4) are standard
	conditions to ensure the smoothness of the mean and covariance function in functional
	data analysis; see \cite{cao2012simultaneous}, \cite{cao2016oracle}, \cite{Zhang2021UnifiedPC} and many others. Assumption (A2) is common in the literature for random design; for example, in \cite{tonycai2011}, \cite{ZW16}. Assumptions (A3) and (A4) include some moment conditions on process $\eta_i(\cdot)$ and stochastic error $\epsilon_{ij}$. Similar moment conditions are also adopted in \cite{ZW16}, \cite{Zhang2021UnifiedPC} and \cite{guo2023sparse}. The range of parameters specified in Assumptions (A1), (A3) and (A4) is given in Assumption (A5). It is worth noting that we do not impose the independence assumption or Gaussian assumption  of FPC scores $\xi_{ik}$'s over $k$ or smoothness of FPCs, compared to that in \cite{cao2012simultaneous}, \cite{cao2016oracle}, \cite{2022lijie} and \cite{zheng2014smooth}, respectively.
	
	{\color{black}
		Define $\mathbf \Sigma=\mathbf\Sigma_1+\mathbf\Sigma_2$,
		where \begin{align*}
			&\mathbf\Sigma_1 = \mathbf{V}^{-1}\E\left\{\frac{\sum_{i=1}^{n}N_i(N_i-1)}{n\bar N^2}\bm B(X)\bm B^\top(X^\prime)G(X,X^\prime)\right\} \mathbf{V}^{-1},\\&
			\mathbf\Sigma_2 = \mathbf{V}^{-1}\E\left\{\frac{1}{\bar N}\bm B(X)\bm B^\top(X)\left(G(X,X)+\sigma^2(X)\right)\right\} \mathbf{V}^{-1}.
		\end{align*}
		Here, $X$ and $X^\prime$ are i.i.d. copies of $X_{ij}$, and $\mathbf{V} = \left\{\E B_{l,p}(X) B_{l^\prime,p}(X)\right\}_{l,l^\prime=1}^{J_n+p}$ is the theoretical inner product matrix.  
		Define a Gaussian process $\mathcal{G}_n(x)$ with mean zero and covariance structure as follows \begin{align*}
			{\text {Cov}}\left(\mathcal{G}_n(x),\mathcal{G}_n(x^\prime)  \right)= \left\|{\mathbf \Sigma}^{1/2}\bm B(x)\right\|_2^{-1}\bm B^\top(x) {\mathbf\Sigma}\bm B(x^\prime)\left\|{\mathbf \Sigma}^{1/2}\bm B(x^\prime)\right\|_2^{-1}.
		\end{align*}

		\begin{theorem}\label{infeasibleSCB}
			Under Assumptions (A1)-(A5), for some Gaussian vector $\bm Z_n\sim N(\bm 0,\mathbf I_{J_n+p})$, as $n\to\infty$, \begin{align*}
				\frac{\sqrt{n}\left\{\hat m(x)-m(x)\right\}}{\left\|\mathbf\Sigma^{1/2}\bm B(x)\right\|_{2}}\overset{d}{=}\frac{\bm B^\top(x)\mathbf\Sigma^{1/2}\bm Z_n}{\left\|\mathbf\Sigma^{1/2}\bm B(x)\right\|_{2}}+R_n(x)\,\, {\text{in}}\,\, C[0,1], 
			\end{align*}
			where $\sup_{x\in[0,1]}|R_n(x)|=\o_p\left(\log^{-1/2}(n)\right)$. We  also have that 
			\begin{align}\label{semi-dense}
				\sup_{t\in\mathbb{R}}\left|\P\left(\sup_{x\in[0,1]}\frac{\sqrt{n}\left|\hat m(x)-m(x)\right|}{\left\|{\mathbf \Sigma}^{1/2}\bm B(x)\right\|_2}\leq t\right)- \P\left(\sup_{x\in[0,1]}\left|\mathcal{G}_n(x)\right| \leq t\right)\right|\to 0.
			\end{align}
		\end{theorem}
		We emphasize that we do not restrict the relationship between $n$ and $N_i$ in Assumptions (A1)-(A5). In other words, $N_i$ can be either fixed, or  diverge to infinity at any rate of the sample size $n$. Hence, 
		Theorem \ref{infeasibleSCB}  provides  an approximation
		of the   process $\sqrt{n}\left\|{\mathbf \Sigma}^{1/2}\bm B(x)\right\|_2^{-1}\left\{\hat m(x)-m(x)\right\}$ by a sequence of zero-mean Gaussian processes $\mathcal{G}_n(\cdot)$ under arbitrary sampling schemes, whether sparse or dense. It can also be construed to assert that, in large samples, the distribution of the process $\hat m(\cdot)-m(\cdot)$ depends on the distribution of the data only via matrix $\mathbf\Sigma$.  
		
		We remark that,  one might hope to obtain a result of the following form in Theorem \ref{infeasibleSCB},
		\begin{align}\label{weakconvergence}
			\frac{\sqrt{n}\left\{\hat m(x)-m(x)\right\}}{\left\|\mathbf\Sigma^{1/2}\bm B(x)\right\|_{2}}\xrightarrow{d} \mathcal G(x)\,\,{\text{in}} \,\,C[0,1].
		\end{align}
		where $\mathcal G(\cdot)$ is some fixed zero-mean Gaussian process. However,   the left hand side of (\ref{weakconvergence}) might not be asymptotically equicontinuous, and so it does not have a limiting distribution.  A similar phenomenon has also been discussed in \cite{belloni2015some}. To address this issue, we provide a sufficient condition to ensure that $\sqrt n\left\{\hat m(\cdot)-m(\cdot)\right\}$ is asymptotically tight in Section \ref{sec2.2}. 
		Besides, it is noted that the constraints on the value of $J_n$ in Assumption (A5)  become increasingly loose as $r$ increases. This is because the existence of higher-order moments reduces the dimensional restrictions for strong approximation in \cite{mies2023sequential}, which aligns with intuition.
		
		\par When the order of $\bar N$ is higher (lower) than the order of $J_n$, one could observe that the matrix $\mathbf\Sigma_2$ converges to $0$ (diverges to infinity) in the spectral norm sense. Therefore, by introducing additional conditions regarding the comparison of orders between $\bar N$ and $J_n$, we could obtain phase transitions of strong approximation in the following  Theorem \ref{infeasibleSCB2}, which shows that the distribution of the process $\hat m(\cdot)-m(\cdot)$ asymptotically depends on the distribution of the data only via matrix $\mathbf\Sigma_2$ in sparse case, and via matrix $\mathbf\Sigma_1$ in dense case. 
		Further define   Gaussian processes $\mathcal{G}_{1n}(x)$, $\mathcal{G}_{2n}(x)$ with mean zero and covariance structure as follows respectively \begin{align*}
			&{\text {Cov}}\left(\mathcal{G}_{1n}(x),\mathcal{G}_{1n}(x^\prime)  \right)= \left\|{\mathbf \Sigma}_1^{1/2}\bm B(x)\right\|_2^{-1}\bm B^\top(x) {\mathbf\Sigma}_1\bm B(x^\prime)\left\|{\mathbf \Sigma}_1^{1/2}\bm B(x^\prime)\right\|_2^{-1},\\&{\text {Cov}}\left(\mathcal{G}_{2n}(x),\mathcal{G}_{2n}(x^\prime)  \right)= \left\|{\mathbf \Sigma}_2^{1/2}\bm B(x)\right\|_2^{-1}\bm B^\top(x) {\mathbf\Sigma}_2\bm B(x^\prime)\left\|{\mathbf \Sigma}_2^{1/2}\bm B(x^\prime)\right\|_2^{-1}.
		\end{align*}
		\begin{theorem}\label{infeasibleSCB2}
			Suppose that Assumptions (A1)-(A4) hold and let $\bm Z_n\sim N(\bm 0,\mathbf I_{J_n+p})$.
			\begin{itemize}
				\item[(1)] Under Assumption (A5) and further assume that $\sum_{i=1}^{n}N_i^2=\O\left(n\bar N^2\right)$, $\bar N\ll J_n\log^{-2}(n)$, as $n\to\infty$,
				\begin{align*}
					\frac{\sqrt{n}\left\{\hat m(x)-m(x)\right\}}{\left\|\mathbf\Sigma_2^{1/2}\bm B(x)\right\|_{2}}\overset{d}{=}\frac{\bm B^\top(x)\mathbf\Sigma_2^{1/2}\bm Z_n}{\left\|\mathbf\Sigma_2^{1/2}\bm B(x)\right\|_{2}}+R_{2n}(x)\,\, {\text{in}}\,\, C[0,1], 
				\end{align*} 
				where $\sup_{x\in[0,1]}|R_{2n}(x)|=\o_p\left(\log^{-1/2}(n)\right)$. Besides, We  have that 
				\begin{align}\label{sparse}
					\sup_{t\in\mathbb{R}}\left|\P\left(\sup_{x\in[0,1]}\frac{\sqrt{n}\left|\hat m(x)-m(x)\right|}{\left\|\mathbf \Sigma_2^{1/2}\bm B(x)\right\|_2}\leq t\right)- \P\left(\sup_{x\in[0,1]}\left|\mathcal{G}_{2n}(x)\right| \leq t\right)\right|\to 0.
				\end{align}
				\item[(2)] Under Assumptions (A1)-(A4) and (A5'), and assume $\bar N\gg J_n\log^2(n)$, as $n\to\infty$, \begin{align*}
					\frac{\sqrt{n}\left\{\hat m(x)-m(x)\right\}}{\left\|\mathbf\Sigma_1^{1/2}\bm B(x)\right\|_{2}}\overset{d}{=}\frac{\bm B^\top(x)\mathbf\Sigma_1^{1/2}\bm Z_n}{\left\|\mathbf\Sigma_1^{1/2}\bm B(x)\right\|_{2}}+R_{1n}(x)\,\, {\text{in}}\,\, C[0,1], 
				\end{align*}
				where $\sup_{x\in[0,1]}|R_{1n}(x)|=\o_p\left(\log^{-1/2}(n)\right)$. Then, We  have that 
				\begin{align}\label{dense}
					\sup_{t\in\mathbb{R}}\left|\P\left(\sup_{x\in[0,1]}\frac{\sqrt{n}\left|\hat m(x)-m(x)\right|}{\left\|\mathbf \Sigma_1^{1/2}\bm B(x)\right\|_2}\leq t\right)- \P\left(\sup_{x\in[0,1]}\left|\mathcal{G}_{1n}(x)\right| \leq t\right)\right|\to 0.
				\end{align}
			\end{itemize}
			
		\end{theorem}

		\par 
		
		Recall that \cite{ZW16} establishes the critical boundary between sparse, semi-dense, and dense by comparing the bias  and variance term (lower-order, same-order, higher-order),  
		when discussing the phase transition in uniform convergence and local asymptotic normality. 
		Therefore, the key to distinguishing between sparse and dense lies in two factors: whether the bias term is asymptotically negligible and whether the randomness introduced by measurement error affects the asymptotic variance, which  involves the   relationship among $n$, $\bar N$  and the non-parametric smoothing parameter. 
		
		In our discussion of phase transitions in simultaneous inference for the mean function, an essential requirement is that the bias term must be asymptotically negligible. If this condition is not met, conducting simultaneous statistical inference becomes challenging. This is because the bias term in the B-spline estimator, or other orthogonal series estimators introduced in Section \ref{sec2.3}, does not have an explicit form and cannot be consistently estimated. To satisfy this requirement, a lower bound for $J_n$ in relation to $n$ and $\bar N$ is necessary. We further note that the phase transition in simultaneous inference primarily depends on the behavior of the variance component $\mathbf\Sigma_2$, which requires comparing the orders of $J_n$ and $\bar N$. However, in existing literature, phase transition results are usually described in terms of the relationship between $n$ and $\bar N$, as it provides a more intuitive understanding. Following this convention, we predetermine the relationship among $J_n$, $n$, and $\bar N$ and then translate the relationship between $J_n$ and $\bar N$ into one between $n$ and $\bar N$.
		
		To achieve the optimal convergence rate discussed in \cite{tonycai2011}, we set $J_n$ close to its lower bound as outlined in Assumption (A5). This allows us to attain the optimal convergence rate, up to a logarithmic term, and establish   phase transitions that vary depending on the smoothness of the mean function. Notably, when $q^\ast=2$, our findings align with the phase transition of uniform convergence in \cite{ZW16}, differing only by a logarithmic term.
		
		\begin{corollary}\label{corollary0}
			Suppose that Assumptions (A1)-(A4) hold.  Further assume that  $\sum_{i=1}^{n}N_i^{2s/r}=\O\left(n\bar N^{2s/r}\right)$ for each $s=1,2,\ldots,r$. For any fixed $b> 1/(2q^\ast)$,
			\begin{itemize}
				\item[(1)]  Under Assumption (A5), when $\bar N\ll  n^{1/(2q^\ast)}\log^{b-(2q^\ast+1)/q^\ast} (n)$, and $J_n\asymp (n\bar N)^{1/(2q^\ast+1)} \log^{ 2bq^\ast /(2q^\ast+1)}(n)$, (\ref{sparse}) holds, referred as ``sparse''. 
				\item[(2)]  Under Assumption (A5), when $n^{1/(2q^\ast)}\log^{b-(2q^\ast+1)/q^\ast} (n)\lesssim \bar N\lesssim n^{1/(2q^\ast)}\\\log^{b+2} (n)$, and $J_n\asymp n^{1/(2q^\ast)} \log^{b}(n)$, (\ref{semi-dense}) holds, referred as ``semi-dense''.
				\item[(3)] Under Assumption (A5'), when $\bar N\gg n^{\frac{1}{2q^\ast}}\log^{b+2} n$,    $J_n\gg  n^{\frac{1}{2q^\ast}}\log^{\frac{1}{2q^\ast}}(n)$, and $J_n\ll \bar N\log^{-2}(n)$, (\ref{dense}) holds, referred as ``dense''.
			\end{itemize}
		\end{corollary}

		\subsection{The oracle property}\label{sec2.2}

		In the following, we provide a sufficient condition, known as the oracle property, to ensure the tightness of the process on the left side of  (\ref{weakconvergence}). In functional data analysis, we often say an estimator enjoys oracle efficiency when it is asymptotically indistinguishable from the estimator derived from completely observed trajectories up to order $\sqrt{n}$.
		To investigate the oracle property of the proposed B-spline estimator, we introduce the following assumptions.
		\begin{itemize}
			\item[(B1)] The rescaled FPCs satisfies $\sum_{k=1}^{\infty}\left\|\phi_k(x)\right\|_{\infty}<\infty$, $\sum_{k=1}^{\infty}\left\|\phi_k(x)\right\|_{0,\varpi}<\infty$ for some small $\varpi\in(0,1]$. 
			\item[(B2)] Suppose that $p\geq q^\ast$, $J_n^{-q^\ast}n^{1/2}=\o(1)$,  $n^{-1}\bar N^{-1}J_n^2\log(n)=\o(1)$ and $\bar N\gg J_n\log(n)$.
		\end{itemize}
		Assumption (B1) is a common condition; see \cite{cao2012simultaneous}, \cite{wang2020simultaneous} for instance, ensuring the  boundedness and H\"{o}lder continuity of FPCs.  The range of parameters is involved in Assumption (B2).  
		\begin{theorem}\label{oracle-theorem}
			Under Assumptions (A1)-(A4) and (B1)-(B2), as $n\to\infty$, we have \begin{align*}
				\sup_{x\in[0,1]}\sqrt{n}\left|\hat m(x)-\frac{1}{n}\sum_{i=1}^{n}\frac{N_i}{\bar N}\eta_i(x)\right|=\o_p(1).
			\end{align*}
			If $r=\lim_{n\to\infty}n^{-1}\bar N^{-2}\sum_{i=1}^{n}N_i^2$ and $ \sum_{i=1}^{n}N_i^{2+\delta}=\O\left(n\bar N^{2+\delta}\right)$ for some small $\delta>0$, we derive that $
			\sqrt{n}\left(\hat m(x)-m(x)\right)\xrightarrow{d} \sqrt r\mathcal{G}_1(x)$,
			where $\mathcal{G}_1(x)$ is a mean-zero Gaussian process with covariance function ${\text{Cov}}\left(\mathcal{G}_1(x),\mathcal{G}_1(x^\prime) \right)=G(x,x^\prime)$, and the weak convergence is in the topology of $C[0,1]$. 
		\end{theorem}
		\par For regular fixed design ($N_i\equiv  N$ and $X_{ij}=j/N$), one needs $J_n^{-q^\ast}n^{ 1/2}=\o(1)$, and $J_n N^{-1}\log(n)=\o(1)$  to ensure the oracle property holds, as demonstrated in \cite{cao2012simultaneous}. These conditions   match the parameter ranges required in Theorem \ref{oracle-theorem}. 
		However, when $J_n\gg N$,  the design matrix may not be full rank, which could result in the spline regression  in (\ref{DEF:mhat}) having no solution. Therefore, discussing the sparse  scenario under regular designs is challenging.
		
		\par To delve deeper into our findings, it's crucial to integrate the insights from Theorems \ref{infeasibleSCB}, \ref{infeasibleSCB2}, \ref{oracle-theorem}, and Corollary \ref{corollary0}. Let's start by considering two extreme scenarios. In the first case, imagine each curve has only a single observation point, i.e., $N_i \equiv 1$. Under these circumstances, our model (\ref{model2}) simplifies to 
		\begin{align*}
			Y_{i} =m(X_{i})+\sum_{k=1}^{\infty} \xi_{ik}\phi_k(X_{i})+\sigma(X_{i})\epsilon_{i}, \quad 1\leq i\leq n,
		\end{align*}
		which resembles a nonparametric regression with independent error and the variance function $G(x,x)+\sigma^2(x)$. If $\phi_k \equiv 0$ for all $k \in \mathbb{N}$, then our proposed SCBs align with the results in \cite{belloni2015some}. Conversely, in the scenario of complete trajectory collection, i.e., $N_i \equiv \infty$, the inference about the mean function essentially applies the Central Limit Theorem in $C[0,1]$.
		
		For sparse cases like those in \cite{zheng2014smooth}, where $N_i$'s are expectation-finite random variables, they constructed SCBs with a rate  $\mathcal{O}_p((nh/\log h^{-1})^{-1/2})$, with bandwidth $h$. Our SCBs exhibit a similar rate $\mathcal{O}_p((n/J_n\log n)^{-1/2})$, aligning with \cite{zheng2014smooth} due to $J_n \asymp h^{-1}$ and Assumption (A5). 
		The intermediate regime ``semi-dense'', situated between the sparse and dense scenarios,  is characterized by the combined influence of variation of trajectories and errors, aligning with the findings presented in \cite{berger2023dense}.
		In semi-dense and dense cases, our convergence rate is $\mathcal{O}_p((n/\log n)^{-1/2})$, identical to that in \cite{ZW16}. In the ultra-dense scenario, where our estimator achieves oracle efficiency, the convergence rate of our SCBs is $\mathcal{O}_p(n^{-1/2})$, matching the results in \cite{cao2012simultaneous}.
		
		The phase transition we discuss hinges on the trade-off between $\Var(Y_{ij})$ and $\Cov(Y_{ij},Y_{ij^\prime})$.
		This is evident in the relationship between $\Sigma_1$ and $\Sigma_2$: $\Sigma_1 \ll \Sigma_2$ in sparse cases, $\Sigma_1 \asymp \Sigma_2$ up to a logarithmic term in semi-dense cases, and $\Sigma_1 \gg \Sigma_2$ in dense cases. Figure \ref{fig:spasetodense} visually illustrates our developed boundary, providing an intuitive understanding of these concepts.

		\begin{figure}
			\centering
			\includegraphics[width=12cm]{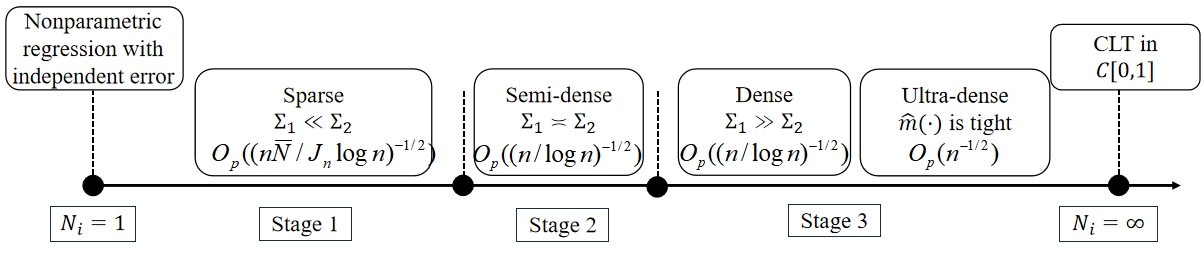}
			\caption{The schema of phase transition phenomena from sparse to dense.}
			\label{fig:spasetodense}
		\end{figure}
		
		\section{Extension to orthogonal  series estimators}\label{sec2.3}
		Let $\{\varphi_l(x)\}_{l=1}^{J_n}$ be some orthogonal basis functions satisfying $\left\|\varphi_l\right\|_{L^2}= 1$. 
		The mean function $m(x)$ can be  also estimated by solving the following orthogonal series regression, defined as:
		\begin{equation}  \label{series}
			\hat{m}(x)=\mathop{\arg\min}
			\limits_{\bm b\in\mathbb R^{J_n}}
			\sum_{i=1}^n\sum_{j=1}^{N_i}  \left\{Y_{ij}-\bm b^\top\bm\varphi
			(X_{ij})\right\}^2,
		\end{equation}
		in which
		$\bm\varphi(x)=\bm\varphi_n(x)=\left(\varphi_1(x),\ldots,\varphi_{J_n}(x)\right)^\top$.   The solution to (\ref{series}) is $\hat m(x)=\bm \varphi^\top (x)\hat{\bm\vartheta}$, where \begin{align*}
			&\hat{\bm\vartheta} =\hat{ \mathbf{Q}}^{-1}\left\{\frac{1}{n\bar N}\sum_{i=1}^{n}\sum_{j=1}^{N_i} \bm\varphi(X_{ij})Y_{ij}\right\}, \quad
			\hat{\mathbf{Q}}=\frac{1}{n\bar N}\sum_{i=1}^{n}\sum_{j=1}^{N_i} \bm\varphi(X_{ij}) \bm\varphi^\top(X_{ij}).
		\end{align*}
		\par The properties of the estimator $\hat m(\cdot)$ depend on the characteristics of the adopted basis. 
		To illustrate the approximation power of the basis, which has impact on the
		convergence rate, we introduce the following two quantities.
		\begin{align*}
			&\kappa^{\bm\varphi}_{1,\infty}=\sup_{m \in\mathcal H_{q,\nu}[0,1]}\inf_{\bm b\in\mathbb R^{J_n}}\left\|m(\cdot)-\bm b^\top \bm\varphi(\cdot) \right\|_{\infty},\\& \kappa^{\bm\varphi}_{2,\infty}=\sup_{G \in\mathcal H_{s,\beta}[0,1]^2}\inf_{\mathbf A \in\mathbb R^{ J_n \times  J_n  }}\left\|G(\cdot,\star)- \bm\varphi^\top(\cdot)\mathbf A \bm\varphi(\star) \right\|_{\infty}.
		\end{align*}
		Another vital feature of the basis associated with  uniform asymptotic behaviors of $\hat m(\cdot)$ is the linear operators $\Pi^{\bm\varphi}_1(\cdot)$ and $\Pi^{\bm\varphi}_{2}(\cdot)$ defined below, 
		\begin{align*}
			&\Pi_1^{\bm\varphi}(g)=\bm\varphi^\top(\cdot) \mathbf Q^{-1}  \E\left\{\bm\varphi(X)g(X)\right\},\\
			&\Pi_2^{\bm\varphi}(h)=\bm\varphi^\top(\cdot)  \mathbf Q^{-1}  \E\left\{\bm\varphi(X)\bm\varphi^\top(X^\prime)h(X,X^\prime)\right\}\mathbf Q^{-1}\bm\varphi(\star), 
		\end{align*}
		for any univariate function $g:[0,1]\to\mathbb R$ with $\left\|g\right\|_{\infty}<\infty$ and bivariate function $h:[0,1]^2\to\mathbb R$ with $\left\|h\right\|_{\infty}<\infty$. Define the $L^{\infty}$ operator norm of $\Pi_1^{\bm\varphi}$ and $\Pi_2^{\bm\varphi}$ by \begin{align*}
			\left\|\Pi_1^{\bm\varphi}\right\|_{L^\infty}=\sup_{\left\|g\right\|_{\infty}<\infty}\left\|\Pi_1^{\bm\varphi}(g)\right\|_{\infty}\left\|g\right\|_{\infty}^{-1},\quad \left\|\Pi_2^{\bm\varphi}\right\|_{L^\infty}=\sup_{\left\|h\right\|_{\infty}<\infty}\left\|\Pi_2^{\bm\varphi}(h)\right\|_{\infty}\left\|h\right\|_{\infty}^{-1}.
		\end{align*}
		Below are some  examples of orthogonal basis and
		their relevant features.
		\par 
		{\textbf{Example 1}}: The Fourier basis functions are defined by $\varphi_1(x)=1$, $\varphi_{2k}(x)=\cos(2k\pi x)$ and  $\varphi_{2k+1}(x)=\sin(2k\pi x)$ for $k\in\mathbb Z_+$,   forming an orthonormal basis of $\mathcal L^2[0,1]$. According to discussions in section 3.1 of \cite{belloni2015some}, $\kappa_{1,\infty}^{\varphi}\lesssim J_n^{-q-\nu}$ and $\kappa_{2,\infty}^{\varphi}\lesssim J_n^{-s-\beta}$  if $m(\cdot)$ and $G(\cdot,\cdot)$ can be extended to  periodic functions; see   \cite{chen2007large} and Corollary 2.4 in Chapter 7 of \cite{devore1993constructive} for reference. 
		From  Example 3.7 of \cite{belloni2015some},
		$\left\|\Pi_1^\varphi\right\|_{L^\infty}\lesssim \log(J_n)$ when  the density of $X$  is the uniform distribution; see also \cite{zygmund2002trigonometric}. 

		\par {\textbf{Example 2}}: Cohen-Deubechies-Vial (CDV) wavelet series of order $s_0$, described in Example 3.4 of \cite{belloni2015some}, form  an orthonormal basis of $\mathcal L^2[0,1]$. See \cite{cohen1993wavelets} for more details on CDV wavelet. According to  Example 3.9 and Section 3.1 of \cite{belloni2015some}, Example 1 of \cite{quan2022optimal}, and Theorem 5.1 of \cite{chen2015optimal}, $\kappa_{1,\infty}^{\varphi}\lesssim J_n^{-q-\nu}$, $\kappa_{2,\infty}^{\varphi}\lesssim J_n^{-s-\beta}$ if the order $s_0\geq \max\{q+\nu,s+\beta\}$, and $\left\|\Pi_1^\varphi\right\|_{L^\infty}\lesssim 1$ if the density of $X$ is bounded away from zero and infinity.
		
		\par {\textbf{Example 3}}: Legendre polynomial series is the orthonormalizd polynomial series $(1,x,x^2,\ldots)$
		with respect to the Lebesgue measure on $[0, 1]$, i.e., $(\varphi_1,\varphi_2,\varphi_3,\ldots) = (1,\sqrt 3 x,\sqrt{5/4}(3x^2-1),\ldots)$. From (3.7) in \cite{belloni2015some} and Theorem 6.2 in Chapter 7 of \cite{devore1993constructive}, one obtains $\kappa^{\bm\varphi}_{1,\infty}\lesssim J_n^{-q-\nu}$ and $\kappa^{\bm\varphi}_{2,\infty}\lesssim J_n^{-s-\beta}$.
		
		Some technical assumptions are imposed for theoretical development. 
		\begin{itemize}
			\item[(C1)] The covariance function $G(\cdot,\cdot)\in\mathcal{H}_{s,\beta}[0,1]^2$ for some integar $s\geq 0$ and some $\beta\in (0,1]$.
			
			\item[(C2)]  Assume that
			$\left\|\varphi_l\right\|_{\infty}\lesssim  l^{\rho}$ and  $\left\|\varphi_l^\prime\right\|_{\infty}\lesssim  l^{\rho+a}$ for some non-negative real number $\rho,a\geq 0$. 
			\item[(C3)] Assume that $J_n^{2\rho+1}n^{-1/2}\bar N^{-1/2}\log(n)=\o(1)$, $\inf_{x\in[0,1]}\left\|\bm\varphi(x)\right\|_2^2\gtrsim J_n$,  $\kappa^{\bm\varphi}_{1,\infty}=\O\left(J_n^{-q^\ast}\right)$, and
			$\kappa_{2,\infty}^{\bm\varphi} \left\|\Pi_2^{\bm\varphi}\right\|_{L^\infty}=\o\left(1\right)$.  When $\bar N\lesssim J_n$, 
			\begin{align*}
				\left\|\Pi_1^{\bm\varphi}\right\|_{L^\infty}J_n^{-q^\ast-1/2}n^{1/2}\bar N^{1/2}\log^{1/2 }(n)=\o(1).
			\end{align*}
			when $\bar N\gg J_n$,
			$\left\|\Pi_1^{\bm\varphi}\right\|_{L^\infty}J_n^{-q^\ast}n^{1/2}\log^{1/2}(n)=\o(1).
			$\\
			For  $r=\min\{r_1,r_2\}$, when $\bar N\lesssim J_n^{2\rho+1}$, 
			\begin{align*}
				J_n^{2\rho +2-2/r} n^{-1+2/r}\bar N \log^2(n)\left\{\frac{1}{n\bar N^2}\sum_{i=1}^{n}N_i^2  \right\}=\o(1).
			\end{align*}
			when $\bar N\gg J_n^{2\rho+1}$,
			\begin{align*}
				J_n^{4\rho +3-2/r} n^{-1+2/r}  \log^2(n)\left\{\frac{1}{n\bar N^2}\sum_{i=1}^{n}N_i^2  \right\}=\o(1).
			\end{align*}
			
			\item[(C3')] Let $r=r_2$ in (C3). 
		\end{itemize}
		Assumption (C2) is mild and can be verified for many frequently-used basis functions, for instance, $\rho=0,a=1$ for the Fourier basis, and $\rho=1/2,a=2$ for the normalized Legendre basis; see Page 66 of  \cite{hardle1990applied}, and Pages 17-18 of the Supplement of \cite{cui2022simultaneous}. We note that the condition $\left\|\varphi_l\right\|_{\infty}\lesssim l^{\rho}$ is imposed to derive $\sup_{x\in[0,1]}\left\|\bm\varphi(x)\right\|_2\lesssim J_n^{\rho+1/2}$, and the CDV series satisfies $\sup_{x\in[0,1]}\left\|\bm\varphi(x)\right\|_2\lesssim J_n^{\rho+1/2}$ with $\rho=0$.  Besides, As shown in the  proofs of Lemmas \ref{LemmaC1} and \ref{LemmaC2} in the supplemental material, one could bound $\left\|\Pi_1^{\bm\varphi}\right\|_{L^\infty}\lesssim J_n^{\rho+1/2}$ and $\left\|\Pi_2^{\bm\varphi}\right\|_{L^\infty}\lesssim J_n^{2\rho+1}$ under Assumption (C2), which are not sharp in many cases, though universally  applicable. Thus, it is sufficient to suppose that $\kappa^{\bm\varphi}_{2,\infty}=\O\left(J_n^{-s-\beta}\right)$   
		with 
		$s+\beta>2\rho+1$ to ensure $\kappa_{2,\infty}^{\bm\varphi} \left\|\Pi_2^{\bm\varphi}\right\|_{L^\infty}=\o\left(1\right)$. Additionally, elementary calculations show that $\inf_{x\in[0,1]}\left\|\bm\varphi(x)\right\|_2^2\gtrsim J_n$ in Assumption (C3) holds at least for the Fourier basis.
		
		
		\par Define $\mathbf \Omega=\mathbf\Omega_1+\mathbf\Omega_2$,
		where \begin{align*}
			&\mathbf\Omega_1 = \mathbf{Q}^{-1}\E\left\{\frac{\sum_{i=1}^{n}N_i(N_i-1)}{n\bar N^2}\bm \varphi(X)\bm \varphi^\top(X^\prime)G(X,X^\prime)\right\} \mathbf{Q}^{-1},\\&
			\mathbf\Omega_2 = \mathbf{Q}^{-1}\E\left\{\frac{1}{\bar N}\bm \varphi(X)\bm \varphi^\top(X)\left(G(X,X)+\sigma^2(X)\right)\right\} \mathbf{Q}^{-1}.
		\end{align*}
		Here, $X$ and $X^\prime$ are i.i.d. copies of $X_{ij}$.
		Define a Gaussian process $\Xi_n(x)$ with mean zero and covariance structure as follows \begin{align*}
			{\text {Cov}}\left(\Xi_n(x),\Xi_n(x^\prime)  \right)= \left\|{\mathbf \Omega}^{1/2}\bm \varphi(x)\right\|_2^{-1}\bm \varphi^\top(x) {\mathbf\Omega}\bm \varphi(x^\prime)\left\|{\mathbf \Omega}^{1/2}\bm \varphi(x^\prime)\right\|_2^{-1}.
		\end{align*}
		Further define   Gaussian processes $\Xi_{1n}(x)$, $\Xi_{2n}(x)$ with mean zero and covariance structure as follows, respectively \begin{align*}
			&{\text {Cov}}\left(\Xi_{1n}(x),\Xi_{1n}(x^\prime)  \right)= \left\|{\mathbf \Omega}_1^{1/2}\bm \varphi(x)\right\|_2^{-1}\bm \varphi^\top(x) {\mathbf\Omega}_1\bm \varphi(x^\prime)\left\|{\mathbf \Omega}_1^{1/2}\bm \varphi(x^\prime)\right\|_2^{-1},\\
			&{\text {Cov}}\left(\Xi_{2n}(x),\Xi_{2n}(x^\prime)  \right)= \left\|{\mathbf \Omega}_2^{1/2}\bm \varphi(x)\right\|_2^{-1}\bm \varphi^\top(x) {\mathbf\Omega}_2\bm \varphi(x^\prime)\left\|{\mathbf \Omega}_2^{1/2}\bm \varphi(x^\prime)\right\|_2^{-1}.
		\end{align*}
		To mimic Theorems \ref{infeasibleSCB} and \ref{infeasibleSCB2}, we derive a unified Gaussian approximation and corresponding phase transitions for orthogonal series estimators.  
		\begin{theorem}\label{infeasibleSCB_general}
			Under Assumptions (A1)-(A4), (C1)-(C3), for some $\bm Z_n\sim N(\bm 0,\mathbf I_{J_n})$, as $n\to\infty$, \begin{align*}
				\frac{\sqrt{n}\left\{\hat m(x)-m(x)\right\}}{\left\|\mathbf\Omega^{1/2}\bm \varphi(x)\right\|_{2}}\overset{d}{=}\frac{\bm \varphi^\top(x)\mathbf\Omega^{1/2}\bm Z_n}{\left\|\mathbf\Omega^{1/2}\bm \varphi(x)\right\|_{2}}+R_n^{\dagger}(x)\,\, {\text{in}}\,\, C[0,1], 
			\end{align*}
			where $\sup_{x\in[0,1]}|R^{\dagger}_n(x)|=\o_p\left(\log^{-1/2}(n)\right)$. We  also have that 
			\begin{align}\label{general-semidense}
				\sup_{t\in\mathbb{R}}\left|\P\left(\sup_{x\in[0,1]}\frac{\sqrt{n}\left|\hat m(x)-m(x)\right|}{\left\|{\mathbf \Omega}^{1/2}\bm \varphi(x)\right\|_2}\leq t\right)- \P\left(\sup_{x\in[0,1]}\left|\Xi_n(x)\right| \leq t\right)\right|\to 0.
			\end{align}
		\end{theorem}
		\begin{theorem}\label{infeasibleSCB2_general}
			Suppose that Assumptions (A1)-(A4), (C1),(C2) hold and let  $\bm Z_n\sim N(\bm 0,\mathbf I_{J_n})$.
			\begin{itemize}
				\item[(1)]  Under  (C3) and further assume $\sum_{i=1}^{n}N_i^2=\O\left(n\bar N^2\right)$ and $\bar N\ll \\J_n\log^{-2}(n)$, as $n\to\infty$, \begin{align*}
					\frac{\sqrt{n}\left\{\hat m(x)-m(x)\right\}}{\left\|\mathbf\Omega_2^{1/2}\bm \varphi(x)\right\|_{2}}\overset{d}{=}\frac{\bm \varphi^\top(x)\mathbf\Omega_2^{1/2}\bm Z_n}{\left\|\mathbf\Omega_2^{1/2}\bm \varphi(x)\right\|_{2}}+R_{2n}^{\dagger}(x)\,\, {\text{in}}\,\, C[0,1], 
				\end{align*}
				where $\sup_{x\in[0,1]}|R_{2n}^{\dagger}(x)|=\o_p\left(\log^{-1/2}(n)\right)$. We  also have that 
				\begin{align}\label{general-sparse}
					\sup_{t\in\mathbb{R}}\left|\P\left(\sup_{x\in[0,1]}\frac{\sqrt{n}\left|\hat m(x)-m(x)\right|}{\left\|\mathbf \Omega_2^{1/2}\bm \varphi(x)\right\|_2}\leq t\right)- \P\left(\sup_{x\in[0,1]}\left|\Xi_{2n}(x)\right| \leq t\right)\right|\to 0.
				\end{align}
				
				\item[(2)]  Under  (C3') and further assume $\bar N\gg J_n^{2\rho+1}\log^2(n)$, as $n\to\infty$, 
				\begin{align*}
					\frac{\sqrt{n}\left\{\hat m(x)-m(x)\right\}}{\left\|\mathbf\Omega_1^{1/2}\bm \varphi(x)\right\|_{2}}\overset{d}{=}\frac{\bm \varphi^\top(x)\mathbf\Omega_1^{1/2}\bm Z_n}{\left\|\mathbf\Omega_1^{1/2}\bm \varphi(x)\right\|_{2}}+R_{1n}^{\dagger}(x)\,\, {\text{in}}\,\, C[0,1], 
				\end{align*}
				where $\sup_{x\in[0,1]}|R_{1n}^{\dagger}(x)|=\o_p\left(\log^{-1/2}(n)\right)$. We  also have that 
				\begin{align}\label{general-dense}
					\sup_{t\in\mathbb{R}}\left|\P\left(\sup_{x\in[0,1]}\frac{\sqrt{n}\left|\hat m(x)-m(x)\right|}{\left\|\mathbf \Omega_1^{1/2}\bm \varphi(x)\right\|_2}\leq t\right)- \P\left(\sup_{x\in[0,1]}\left|\Xi_{1n}(x)\right| \leq t\right)\right|\to 0.
				\end{align}
			\end{itemize}
			
		\end{theorem}
		It is noted that the constraint on the range of parameter $J_n$ in assumption (C3) is more stringent than that in (A5). 
		These are due to the local compact support property of B-splines, as well as the appealing properties of uniform bound for the bias terms in B-spline estimators, highlighting their advantages over general orthogonal basis.
		\par Inspired by Theorems \ref{infeasibleSCB_general} and \ref{infeasibleSCB2_general}, we state the phase transition as follows. \begin{corollary}
			Suppose that Assumptions (A1)-(A4), (C1)-(C2) hold. Further assume that  $\sum_{i=1}^{n}N_i^2=\O\left(n\bar N^2\right)$ and $\left\|\Pi_1^{\bm\varphi}\right\|_{L^\infty}=\O\left(\log^{\tau}(n)\right)$ for some $\tau>0$. For any fixed $b>(2\tau+1)/(2 q^\ast)$,
			\begin{itemize}
				\item[(1)] Under Assumption (C3), when $\bar N\ll  n^{1/(2q^\ast) }\log^{b-(2q^\ast+1)/q^\ast}(n)$, and $J_n\asymp \left(n\bar N\right)^{1/(2q^\ast+1)} \log^{2bq^\ast /(2q^\ast+1)}(n)$, (\ref{general-sparse}) holds, referred as ``sparse''.
				\item[(2)]Under Assumption (C3), when $  n^{1/(2q^\ast) }\log^{b-(2q^\ast+1)/q^\ast}(n) \lesssim \bar N\lesssim \\n^{(2\rho+1)/(2q^\ast)}\log^{(2\rho+1)b+2}(n)$, and  $J_n\asymp  n^{1/(2q^\ast)}\log^{b}(n)$, (\ref{general-semidense}) holds, referred as ``semi-dense''.
				\item[(3)]Under Assumption (C3'), when $\bar N\gg  n^{(2\rho+1)/(2q^\ast) }\log^{(2\rho+1)b+2}(n)$, and  $J_n\gg   n^{1/(2q^\ast)}\log^{b}(n)$, $J_n^{2\rho+1}\log^2(n)\ll \bar N$, (\ref{general-dense}) holds, referred as ``dense''.
			\end{itemize}
		\end{corollary}

		\section{Implementation}\label{sec3}
		In this section, we focus on the estimation of the covariance structure of  the proposed B-spline estimator. Since it is not difficult to extend to  that of the orthogonal series estimator developed in Section \ref{sec2.3}, details are omitted to save space. 
		\par 
		Denote by $\hat U_{ij}=Y_{ij}-\hat m(X_{ij})$. The matrix $\mathbf\Sigma$   can be estimated by $\hat{\mathbf \Sigma}=\hat{\mathbf \Sigma}_1+\hat{\mathbf \Sigma}_2$, where
		\begin{equation}\label{Sigma}
			\begin{aligned}
				& \hat {\mathbf \Sigma}_1=\hat{\mathbf V}^{-1}\left\{\frac{1}{n\bar N^2}\sum_{i=1}^{n}\sum_{j\neq j^\prime}^{N_i}\bm B(X_{ij})\bm B^\top (X_{ij^\prime})\hat U_{ij}\hat U_{ij^\prime}\right\}\hat{\mathbf V}^{-1},\\
				&\hat {\mathbf \Sigma}_2=\hat{\mathbf V}^{-1}\left\{\frac{1}{n\bar N^2}\sum_{i=1}^{n}\sum_{j=1}^{N_i}\bm B(X_{ij})\bm B^\top(X_{ij})\hat U_{ij}^2\right\}\hat{\mathbf V}^{-1}.
			\end{aligned}
		\end{equation}
		Compared with the covariance estimation in \cite{yao2005functional}, \cite{cao2012simultaneous}, \cite{cao2016oracle}, \cite{ZW16} and many others, we emphasize that  the proposed estimator $\hat{\mathbf\Sigma}$ of the covariance structure is   guaranteed
		to be positive semi-definite,  thus no additional procedures are employed to adjust the estimated covariance structure.  
		

		To proceed, it suffices to obtain the quantile of the sup-norm of $\mathcal{G}_n(\cdot)$,
		denote by $Q_{1-\alpha}$, to construct feasible SCB for the mean function from sparse to dense.
			It would be challenging to directly derive the distribution $\sup_{x\in[0,1]}\left|\mathcal{G}_n(x)\right|$ analytically and calculate its quantile $Q_{1-\alpha}$. To obtain the estimated quantile, it is vital to generate Gaussian process, denoted by $\zeta_n(\cdot)$, whose covariance structure mimics that of $\mathcal{G}_n(\cdot)$ by the plug-in principle. 
			The following  describes how to efficiently generate a sequence of  random processes $\{\zeta_{b,n}(\cdot)\}_{b=1}^{B}$, which are i.i.d. copies of $\zeta_n(\cdot)$.
			Given the data $\{X_{ij},Y_{ij}\}_{i=1,j=1}^{n,N_i}$, one generates i.i.d. $(J_n+p)$-dimensional Gaussian random vectors $\{\bm Z_{b,n}\}_{b=1}^{B}$ with mean zero and covariance matrix $\hat{\mathbf \Sigma}$ and let $\zeta_{b,n}(\cdot)=\left\|\hat{\mathbf \Sigma}^{1/2}\bm B(\cdot)\right\|_2^{-1}\bm B^\top(\cdot)\bm Z_{b,n}$.
			Thus, we have
			\begin{align*}
				{\text {Cov}}\left(\zeta_n(x),\zeta_n(x^\prime)  \right)= \left\|\hat{\mathbf \Sigma}^{1/2}\bm B(x)\right\|_2^{-1}\bm B^\top(x) \hat{\mathbf\Sigma}\bm B(x^\prime)\left\|\hat{\mathbf \Sigma}^{1/2}\bm B(x^\prime)\right\|_2^{-1}.
			\end{align*}

			The following theorem shows the uniform consistency of the estimated covariance structure and the asymptotic standard deviation function of process $\sqrt n\left\{\hat m(\cdot)-m(\cdot)\right\}$. 
			\begin{itemize}
				\item[(D1)] The parameters $r_1,r_2\geq 4$ in Assumption (A3), (A4).
			\end{itemize}
			
			\begin{theorem}\label{Consistency}
				Assume that $\sum_{i=1}^{n}N_i(N_i-1)=\O\left(n\bar N^2\right)$, $\sum_{i=1}^{n}N_i(N_i-1)^2=\O\left(n\bar N^3\right)$, and  $\sum_{i=1}^{n}N_i(N_i-1)(N_i-2)(N_i-3)=\O\left(n\bar N^4\right)$. Under Assumptions (A1)-(A5), (D1) and $n^{-1/2}\bar N^{-1/2}J_n^{2} \log^{1/2}(n)=\o(1)$,  as $n\to\infty$, 
				\begin{align*}
					&\sup_{x,x^\prime\in[0,1]}\left| {\text {Cov}}\left(\zeta_n(x),\zeta_n(x^\prime)  \right)- {\text {Cov}}\left(\mathcal G_n(x),\mathcal G_n(x^\prime)  \right)  \right|=\o_p\left(1\right),\\
					& \sup_{x\in[0,1]}\left| \left\|\hat{\mathbf \Sigma}^{1/2}\bm B(x)\right\|_2-\left\|\mathbf \Sigma^{1/2}\bm B(x)\right\|_2\right|=\o_p\left(1\right).
				\end{align*}
			\end{theorem}
			
			Denote by $\hat Q_{1-\alpha}$ the upper-$\alpha$ quantile of the sup-norm of $\zeta_n(\cdot)$. Then, the asymptotically correct SCB of confidence level $1-\alpha$ for the mean function is defined by 
			\begin{align*}
				\left(\hat m(x)- \frac{\hat Q_{1-\alpha}}{\sqrt{n}}\left\|\hat{\mathbf \Sigma}^{1/2}\bm B(x)\right\|_2 , \hat m(x)+\frac{\hat Q_{1-\alpha}}{\sqrt{n}}\left\|\hat{\mathbf \Sigma}^{1/2}\bm B(x)\right\|_2\right),\quad x\in[0,1].
			\end{align*}


			\section{Numerical simulations}
			In this section, numerical simulation studies are conducted to assess the finite-sample performance of the proposed SCB via B-splines. The data are generated from model (\ref{model2}).
			A total of four sampling schemes from sparse to dense are taken into consideration. We  assume that $N_i$  are i.i.d. from a discrete uniform distribution on the set: (1) $\{3,4,5,6\}$, (2) $\{\lfloor 2n^{1/5} \rfloor, \lfloor 2n^{1/5} \rfloor+1, \ldots, \lfloor 4n^{1/5} \rfloor\}$, (3) $\{\lfloor  n^{1/2} \rfloor, \lfloor n^{1/2} \rfloor+1, \ldots, \lfloor 2n^{1/2} \rfloor\}$, and (4) $\{\lfloor  n/4 \rfloor, \lfloor n/4 \rfloor+1, \ldots, \lfloor n/2 \rfloor\}$.

			In each case, we set $m(x)=3/2\sin\{3\pi(x+ 1/2)\} + 2x^3$,  $\lambda_k=2^{1-k}$ for $k=1,\ldots,4$ and $\lambda_k=0$ for $k\geq 5$, $\psi_{2k-1}=\sqrt{2}\sin(2k\pi x)$ and $\psi_{2k}=\sqrt{2}\cos(2k\pi x)$ for $k\in\mathbb{Z}_+$. Besides, $\{\xi_{tk}\}_{t=1,k=1}^{n,\infty}$ are i.i.d random variables.
			We  consider the variance function $\sigma^2(\cdot)$ as in a homoscedastic case with $\sigma(x) = \sigma_{\varepsilon }$ and in a heteroscedastic case with $\sigma(x) = 1.2\sigma_{\varepsilon }\left( 5 - \exp (-x)\right)/ \left( 5 + \exp (-x)\right).$ The noise level is set at $\sigma_{\varepsilon }=0.1$. We also consider different distributions of the FPC scores and the errors, including the standard Normal distribution $N(0, 1)$, the Uniform distribution $U(-\sqrt{3}, \sqrt{3})$, and the standardized Laplace distribution with density $f(x)=2^{-1/2} \exp(\sqrt{2}|x|)$. The number of sample size $n$ is set at $100$, $200$, and $400$. The empirical  replications $B$ in quantile estimation and the Monte Carlo replications are both $500$. 
			\par The number of knots $J_n$ is a fundamental smoothing parameter and a data-driven method proposed in \cite{huang2004identification} that minimizes the BIC value corresponding to $J_{n} \in [0.5(n\bar{N})^{1/6},2(n\bar{N})^{1/4}]$ is what we recommend. Note that this specific range satisfies the assumptions and contains enough candidates without increasing too much computational burden. The BIC value is defined as
			\begin{equation}\label{BIC}
				\mathrm{BIC}\left(J_{n}\right) = \log\left[(n\bar{N})^{-1}\sum_{i = 1}^{n}\sum_{j = 1}^{N_i}\left\{Y_{ij} - \widehat{m}_{J_n}(X_{ij})\right\}^2\right] + \frac{J_n\log(n\bar{N})}{n\bar{N}},
			\end{equation}
			where $\widehat{m}_{J_n}$ is the estimator defined in (\ref{DEF:mhat}) with the number of knots $J_n$.
			
			\par For comparison, we also implement the simultaneous inference procedure developed in \cite{cao2012simultaneous}, denoted by CYT in Tables \ref{n=100}-\ref{n=400}. To illustrate the asymptotic variance decomposition transitioning from sparse to dense, we also report the values of $\|\hat{\mathbf \Sigma}_1\|_{\infty}$ and $\|\hat{\mathbf \Sigma}_2\|_{\infty}$. 
			Tables \ref{n=100}-\ref{n=400} show that in each setting, the empirical coverage rates of the proposed SCBs  are
			close to the nominal confidence level $95\%$. Figures \ref{n=200figure} and \ref{n=400figure} respectively display the estimated mean functions and corresponding $95\%$ SCBs  with $n =200$ and $n= 400$ in different settings, which show the true mean function is fully encompassed within the SCBs and the SCBs narrow as $n$ increases. These strongly reveal a positive confirmation of the asymptotic
			theory. However, it is seen that the  method developed in \cite{cao2012simultaneous}  performs well only in the dense scenario with a large sample size, namely experiments with  $n=400$ in Setting 4. 
			
			\begin{table*}[h]
				\centering
				\caption{Empirical coverage rates of $95\%$  SCBs   for the mean function $m(x)$  with a sample size of  $n=100$.}
				\resizebox{\textwidth}{!}{
					\begin{tabular}{ccllllll}
						\toprule
						&       & \multicolumn{6}{c}{Homoscedastic (Heteroscedastic) error} \\
						\cmidrule{3-8}    $\xi$    & $\epsilon$ & \multicolumn{1}{c}{Normal} & \multicolumn{1}{c}{Uniform} & \multicolumn{1}{c}{Laplace} & \multicolumn{1}{c}{Normal} & \multicolumn{1}{c}{Uniform} & \multicolumn{1}{c}{Laplace} \\
						\midrule
						&       & \multicolumn{3}{c}{Setting 1} & \multicolumn{3}{c}{Setting 2} \\
						\cmidrule{3-8}    \multirow{4}[2]{*}{Normal} & New   & 0.900(0.898) & 0.882(0.894) & 0.898(0.908) & \multicolumn{1}{c}{0.884(0.926)} & \multicolumn{1}{c}{0.902(0.894)} & \multicolumn{1}{c}{0.926(0.940)} \\
						& CYT   & 0.240(0.260) & 0.208(0.252) & 0.260(0.270) & \multicolumn{1}{c}{0.386(0.444)} & \multicolumn{1}{c}{0.406(0.400)} & \multicolumn{1}{c}{0.444(0.438)} \\
						& $\|\hat{\mathbf \Sigma}_1\|_{\infty}$ & 10.23(10.49) & 10.30(10.25) & 10.49(10.21) & \multicolumn{1}{c}{7.24(7.25)} & \multicolumn{1}{c}{7.11(7.09)} & \multicolumn{1}{c}{7.25(7.07)} \\
						& $\|\hat{\mathbf \Sigma}_2\|_{\infty}$ & 34.03(35.05) & 33.99(33.91) & 35.05(33.94) & \multicolumn{1}{c}{15.86(16.09)} & \multicolumn{1}{c}{15.74(15.90)} & \multicolumn{1}{c}{16.09(15.74)} \\
						\midrule
						\multirow{4}[2]{*}{Uniform} & New   & 0.884(0.864) & 0.890(0.884) & 0.864(0.886) & \multicolumn{1}{c}{0.924(0.900)} & \multicolumn{1}{c}{0.882(0.900)} & \multicolumn{1}{c}{0.900(0.914)} \\
						& CYT   & 0.194(0.234) & 0.242(0.200) & 0.234(0.188) & \multicolumn{1}{c}{0.396(0.378)} & \multicolumn{1}{c}{0.418(0.424)} & \multicolumn{1}{c}{0.378(0.400)} \\
						& $\|\hat{\mathbf \Sigma}_1\|_{\infty}$ & 9.24(9.39) & 9.32(9.14) & 9.39(9.12) & \multicolumn{1}{c}{6.63(6.56)} & \multicolumn{1}{c}{6.72(6.41)} & \multicolumn{1}{c}{6.56(6.57)} \\
						& $\|\hat{\mathbf \Sigma}_2\|_{\infty}$ & 33.04(33.27) & 33.23(33.32) & 33.27(33.23) & \multicolumn{1}{c}{15.48(15.47)} & \multicolumn{1}{c}{15.57(15.47)} & \multicolumn{1}{c}{15.47(15.57)} \\
						\midrule
						\multirow{4}[2]{*}{Laplace} & New   & 0.898(0.908) & 0.888(0.890) & 0.908(0.916) & \multicolumn{1}{c}{0.912(0.914)} & \multicolumn{1}{c}{0.910(0.918)} & \multicolumn{1}{c}{0.914(0.922)} \\
						& CYT   & 0.236(0.206) & 0.252(0.246) & 0.206(0.250) & \multicolumn{1}{c}{0.426(0.450)} & \multicolumn{1}{c}{0.418(0.448)} & \multicolumn{1}{c}{0.450(0.442)} \\
						& $\|\hat{\mathbf \Sigma}_1\|_{\infty}$ & 12.16(12.20) & 12.05(12.04) & 12.20(12.65) & \multicolumn{1}{c}{8.17(8.36)} & \multicolumn{1}{c}{8.05(7.95)} & \multicolumn{1}{c}{8.36(8.12)} \\
						& $\|\hat{\mathbf \Sigma}_2\|_{\infty}$ & 35.50(35.93) & 34.53(35.46) & 35.93(35.70) & \multicolumn{1}{c}{16.47(16.61)} & \multicolumn{1}{c}{16.24(16.28)} & \multicolumn{1}{c}{16.61(16.61)} \\
						\midrule
						&       & \multicolumn{3}{c}{Setting 3} & \multicolumn{3}{c}{Setting 4} \\
						\cmidrule{3-8}    \multirow{4}[2]{*}{Normal} & New   & 0.926(0.928) & 0.926(0.928) & 0.928(0.938) & 0.938(0.940) & 0.930(0.944) & 0.940(0.924) \\
						& CYT   & 0.534(0.576) & 0.564(0.586) & 0.576(0.542) & 0.778(0.790) & 0.770(0.780) & 0.790(0.746) \\
						& $\|\hat{\mathbf \Sigma}_1\|_{\infty}$ & 6.33(6.22) & 6.49(6.36) & 6.22(6.19) & 5.67(5.65) & 5.68(5.62) & 5.65(5.67) \\
						& $\|\hat{\mathbf \Sigma}_2\|_{\infty}$ & 9.25(9.09) & 9.22(9.31) & 9.09(9.15) & 3.58(3.57) & 3.60(3.60) & 3.57(3.60) \\
						\midrule
						\multirow{4}[2]{*}{Uniform} & New   & 0.910(0.914) & 0.930(0.890) & 0.914(0.906) & 0.944(0.942) & 0.930(0.934) & 0.942(0.918) \\
						& CYT   & 0.556(0.528) & 0.546(0.538) & 0.528(0.508) & 0.774(0.774) & 0.774(0.780) & 0.774(0.746) \\
						& $\|\hat{\mathbf \Sigma}_1\|_{\infty}$ & 6.04(6.03) & 5.90(6.04) & 6.03(5.94) & 5.48(5.58) & 5.54(5.57) & 5.58(5.47) \\
						& $\|\hat{\mathbf \Sigma}_2\|_{\infty}$ & 9.12(9.10) & 9.12(9.11) & 9.10(9.03) & 3.58(3.55) & 3.56(3.57) & 3.55(3.55) \\
						\midrule
						\multirow{4}[2]{*}{Laplace} & New   & 0.894(0.916) & 0.934(0.946) & 0.916(0.918) & 0.940(0.932) & 0.922(0.922) & 0.932(0.948) \\
						& CYT   & 0.566(0.574) & 0.606(0.592) & 0.574(0.562) & 0.788(0.768) & 0.744(0.782) & 0.768(0.800) \\
						& $\|\hat{\mathbf \Sigma}_1\|_{\infty}$ & 7.02(7.02) & 7.21(7.02) & 7.02(7.07) & 5.91(5.91) & 6.03(5.86) & 5.91(5.97) \\
						& $\|\hat{\mathbf \Sigma}_2\|_{\infty}$ & 9.39(9.37) & 9.36(9.37) & 9.37(9.51) & 3.69(3.64) & 3.67(3.64) & 3.64(3.65) \\
						\bottomrule
					\end{tabular}%
					\label{n=100}%
				}
			\end{table*}%
			
			\begin{table*}[h]
				\centering
				\caption{Empirical coverage rates of $95\%$  SCBs   for the mean function $m(x)$  with a sample size of  $n=200$.}
				\resizebox{\textwidth}{!}{
					\begin{tabular}{ccllllll}
						\toprule
						&       & \multicolumn{6}{c}{Homoscedastic (Heteroscedastic) error} \\
						\cmidrule{3-8}    $\xi$    & $\epsilon$ & \multicolumn{1}{c}{Normal} & \multicolumn{1}{c}{Uniform} & \multicolumn{1}{c}{Laplace} & \multicolumn{1}{c}{Normal} & \multicolumn{1}{c}{Uniform} & \multicolumn{1}{c}{Laplace} \\
						\midrule
						&       & \multicolumn{3}{c}{Setting 1} & \multicolumn{3}{c}{Setting 2} \\
						\cmidrule{3-8}    \multirow{4}[2]{*}{Normal} & New   & 0.912(0.924) & 0.928(0.924) & 0.924(0.942) & \multicolumn{1}{c}{0.934(0.916)} & \multicolumn{1}{c}{0.926(0.916)} & \multicolumn{1}{c}{0.916(0.940)} \\
						& CYT   & 0.192(0.180) & 0.190(0.200) & 0.180(0.192) & \multicolumn{1}{c}{0.466(0.418)} & \multicolumn{1}{c}{0.482(0.440)} & \multicolumn{1}{c}{0.418(0.444)} \\
						& $\|\hat{\mathbf \Sigma}_1\|_{\infty}$ & 8.47(8.52) & 8.47(8.81) & 8.52(8.48) & \multicolumn{1}{c}{5.26(5.43)} & \multicolumn{1}{c}{5.37(5.41)} & \multicolumn{1}{c}{5.43(5.17)} \\
						& $\|\hat{\mathbf \Sigma}_2\|_{\infty}$ & 36.54(36.52) & 36.52(36.96) & 36.52(36.40) & \multicolumn{1}{c}{13.59(13.55)} & \multicolumn{1}{c}{13.51(13.52)} & \multicolumn{1}{c}{13.55(13.49)} \\
						\midrule
						\multirow{4}[2]{*}{Uniform} & New   & 0.908(0.906) & 0.902(0.900) & 0.906(0.910) & \multicolumn{1}{c}{0.914(0.932)} & \multicolumn{1}{c}{0.924(0.934)} & \multicolumn{1}{c}{0.932(0.930)} \\
						& CYT   & 0.164(0.170) & 0.200(0.182) & 0.170(0.168) & \multicolumn{1}{c}{0.432(0.456)} & \multicolumn{1}{c}{0.468(0.470)} & \multicolumn{1}{c}{0.456(0.480)} \\
						& $\|\hat{\mathbf \Sigma}_1\|_{\infty}$ & 7.59(7.60) & 7.82(7.71) & 7.60(7.58) & \multicolumn{1}{c}{4.89(4.94)} & \multicolumn{1}{c}{4.93(4.95)} & \multicolumn{1}{c}{4.94(5.03)} \\
						& $\|\hat{\mathbf \Sigma}_2\|_{\infty}$ & 36.21(36.02) & 36.11(36.09) & 36.02(35.71) & \multicolumn{1}{c}{13.45(13.40)} & \multicolumn{1}{c}{13.49(13.51)} & \multicolumn{1}{c}{13.40(13.44)} \\
						\midrule
						\multirow{4}[2]{*}{Laplace} & New   & 0.932(0.912) & 0.922(0.926) & 0.912(0.930) & \multicolumn{1}{c}{0.954(0.934)} & \multicolumn{1}{c}{0.938(0.928)} & \multicolumn{1}{c}{0.934(0.936)} \\
						& CYT   & 0.210(0.218) & 0.212(0.202) & 0.218(0.218) & \multicolumn{1}{c}{0.496(0.476)} & \multicolumn{1}{c}{0.484(0.446)} & \multicolumn{1}{c}{0.476(0.492)} \\
						& $\|\hat{\mathbf \Sigma}_1\|_{\infty}$ & 10.44(10.37) & 10.53(9.86) & 10.37(10.48) & \multicolumn{1}{c}{6.18(6.08)} & \multicolumn{1}{c}{6.11(6.17)} & \multicolumn{1}{c}{6.08(6.02)} \\
						& $\|\hat{\mathbf \Sigma}_2\|_{\infty}$ & 37.95(38.15) & 38.33(37.71) & 38.15(38.03) & \multicolumn{1}{c}{13.93(13.86)} & \multicolumn{1}{c}{13.96(13.84)} & \multicolumn{1}{c}{13.86(13.62)} \\
						\midrule
						&       & \multicolumn{3}{c}{Setting 3} & \multicolumn{3}{c}{Setting 4} \\
						\cmidrule{3-8}    \multirow{4}[2]{*}{Normal} & New   & 0.946(0.920) & 0.938(0.944) & 0.920(0.906) & 0.938(0.958) & 0.934(0.950) & 0.958(0.942) \\
						& CYT   & 0.618(0.564) & 0.602(0.580) & 0.564(0.620) & 0.874(0.846) & 0.848(0.880) & 0.846(0.850) \\
						& $\|\hat{\mathbf \Sigma}_1\|_{\infty}$ & 4.73(4.74) & 4.76(4.66) & 4.74(4.70) & 4.19(4.20) & 4.17(4.16) & 4.20(4.19) \\
						& $\|\hat{\mathbf \Sigma}_2\|_{\infty}$ & 7.31(7.31) & 7.30(7.25) & 7.31(7.30) & 2.01(2.00) & 2.00(2.00) & 2.00(2.01) \\
						\midrule
						\multirow{4}[2]{*}{Uniform} & New   & 0.928(0.918) & 0.920(0.942) & 0.918(0.940) & 0.950(0.946) & 0.950(0.948) & 0.946(0.926) \\
						& CYT   & 0.614(0.642) & 0.620(0.650) & 0.642(0.604) & 0.862(0.846) & 0.850(0.820) & 0.846(0.844) \\
						& $\|\hat{\mathbf \Sigma}_1\|_{\infty}$ & 4.48(4.47) & 4.51(4.46) & 4.47(4.44) & 4.10(4.07) & 4.10(4.12) & 4.07(4.09) \\
						& $\|\hat{\mathbf \Sigma}_2\|_{\infty}$ & 7.24(7.23) & 7.25(7.22) & 7.23(7.29) & 2.00(2.00) & 2.00(1.99) & 2.00(2.00) \\
						\midrule
						\multirow{4}[2]{*}{Laplace} & New   & 0.918(0.948) & 0.928(0.944) & 0.948(0.934) & 0.946(0.942) & 0.938(0.944) & 0.942(0.966) \\
						& CYT   & 0.628(0.694) & 0.644(0.624) & 0.694(0.608) & 0.836(0.864) & 0.860(0.826) & 0.864(0.890) \\
						& $\|\hat{\mathbf \Sigma}_1\|_{\infty}$ & 5.28(5.14) & 5.07(5.07) & 5.14(5.18) & 4.37(4.32) & 4.32(4.40) & 4.32(4.33) \\
						& $\|\hat{\mathbf \Sigma}_2\|_{\infty}$ & 7.40(7.38) & 7.33(7.42) & 7.38(7.44) & 2.03(2.03) & 2.00(2.02) & 2.03(2.01) \\
						\bottomrule
					\end{tabular}%
				}
				\label{n=200}%
			\end{table*}%
			
			\begin{table*}[h]
				\centering
				\caption{Empirical coverage rates of $95\%$  SCBs   for the mean function $m(x)$  with a sample size of $n=400$.}
				\resizebox{\textwidth}{!}{
					\begin{tabular}{ccllllll}
						\toprule
						&       & \multicolumn{6}{c}{Homoscedastic (Heteroscedastic) error} \\
						\cmidrule{3-8}    $\xi$    & $\epsilon$ & \multicolumn{1}{c}{Normal} & \multicolumn{1}{c}{Uniform} & \multicolumn{1}{c}{Laplace} & \multicolumn{1}{c}{Normal} & \multicolumn{1}{c}{Uniform} & \multicolumn{1}{c}{Laplace} \\
						\midrule
						&       & \multicolumn{3}{c}{Setting 1} & \multicolumn{3}{c}{Setting 2} \\
						\cmidrule{3-8}    \multirow{4}[2]{*}{Normal} & New   & 0.944(0.914) & 0.938(0.918) & 0.914(0.942) & \multicolumn{1}{c}{0.938(0.940)} & \multicolumn{1}{c}{0.926(0.942)} & \multicolumn{1}{c}{0.940(0.944)} \\
						& CYT   & 0.134(0.140) & 0.140(0.114) & 0.140(0.150) & \multicolumn{1}{c}{0.510(0.474)} & \multicolumn{1}{c}{0.508(0.492)} & \multicolumn{1}{c}{0.474(0.456)} \\
						& $\|\hat{\mathbf \Sigma}_1\|_{\infty}$ & 7.05(7.08) & 7.01(7.10) & 7.08(7.16) & \multicolumn{1}{c}{4.22(4.16)} & \multicolumn{1}{c}{4.17(4.21)} & \multicolumn{1}{c}{4.16(4.22)} \\
						& $\|\hat{\mathbf \Sigma}_2\|_{\infty}$ & 39.27(39.12) & 39.37(39.32) & 39.12(39.49) & \multicolumn{1}{c}{11.31(11.32)} & \multicolumn{1}{c}{11.33(11.30)} & \multicolumn{1}{c}{11.32(11.37)} \\
						\midrule
						\multirow{4}[2]{*}{Uniform} & New   & 0.926(0.960) & 0.934(0.934) & 0.960(0.930) & \multicolumn{1}{c}{0.956(0.940)} & \multicolumn{1}{c}{0.912(0.934)} & \multicolumn{1}{c}{0.940(0.950)} \\
						& CYT   & 0.136(0.120) & 0.114(0.138) & 0.120(0.134) & \multicolumn{1}{c}{0.462(0.472)} & \multicolumn{1}{c}{0.444(0.492)} & \multicolumn{1}{c}{0.472(0.518)} \\
						& $\|\hat{\mathbf \Sigma}_1\|_{\infty}$ & 6.26(6.31) & 6.34(6.50) & 6.31(6.21) & \multicolumn{1}{c}{3.98(4.01)} & \multicolumn{1}{c}{4.02(3.92)} & \multicolumn{1}{c}{4.01(3.97)} \\
						& $\|\hat{\mathbf \Sigma}_2\|_{\infty}$ & 38.92(38.94) & 38.66(38.88) & 38.94(38.83) & \multicolumn{1}{c}{11.27(11.28)} & \multicolumn{1}{c}{11.30(11.25)} & \multicolumn{1}{c}{11.28(11.27)} \\
						\midrule
						\multirow{4}[1]{*}{Laplace} & New   & 0.940(0.948) & 0.942(0.924) & 0.948(0.940) & \multicolumn{1}{c}{0.936(0.940)} & \multicolumn{1}{c}{0.932(0.944)} & \multicolumn{1}{c}{0.940(0.936)} \\
						& CYT   & 0.128(0.156) & 0.162(0.140) & 0.156(0.142) & \multicolumn{1}{c}{0.516(0.482)} & \multicolumn{1}{c}{0.472(0.482)} & \multicolumn{1}{c}{0.482(0.480)} \\
						& $\|\hat{\mathbf \Sigma}_1\|_{\infty}$ & 8.38(8.79) & 8.79(8.75) & 8.79(8.64) & \multicolumn{1}{c}{4.65(4.62)} & \multicolumn{1}{c}{4.75(4.70)} & \multicolumn{1}{c}{4.62(4.80)} \\
						& $\|\hat{\mathbf \Sigma}_2\|_{\infty}$ & 40.80(40.77) & 40.59(40.47) & 40.77(40.52) & \multicolumn{1}{c}{11.43(11.42)} & \multicolumn{1}{c}{11.53(11.43)} & \multicolumn{1}{c}{11.42(11.50)}  \\ \midrule
						&       & \multicolumn{3}{c}{Setting 3} & \multicolumn{3}{c}{Setting 4} \\
						\cmidrule{3-8}    \multirow{4}[2]{*}{Normal} & New   & 0.950(0.938) & 0.954(0.948) & 0.938(0.958) & 0.944(0.948) & 0.952(0.948) & 0.948(0.944) \\
						& CYT   & 0.686(0.716) & 0.712(0.688) & 0.716(0.688) & 0.926(0.924) & 0.932(0.924) & 0.924(0.932) \\
						& $\|\hat{\mathbf \Sigma}_1\|_{\infty}$ & 3.83(3.86) & 3.88(3.88) & 3.86(3.85) & 3.48(3.50) & 3.48(3.50) & 3.50(3.50) \\
						& $\|\hat{\mathbf \Sigma}_2\|_{\infty}$ & 5.58(5.61) & 5.59(5.62) & 5.61(5.57) & 1.11(1.11) & 1.11(1.11) & 1.11(1.11) \\
						\midrule
						\multirow{4}[2]{*}{Uniform} & New   & 0.918(0.930) & 0.948(0.930) & 0.930(0.936) & 0.954(0.946) & 0.956(0.950) & 0.946(0.946) \\
						& CYT   & 0.690(0.640) & 0.658(0.676) & 0.640(0.668) & 0.934(0.924) & 0.946(0.930) & 0.924(0.932) \\
						& $\|\hat{\mathbf \Sigma}_1\|_{\infty}$ & 3.71(3.77) & 3.74(3.71) & 3.77(3.74) & 3.44(3.44) & 3.44(3.44) & 3.44(3.44) \\
						& $\|\hat{\mathbf \Sigma}_2\|_{\infty}$ & 5.60(5.61) & 5.58(5.59) & 5.61(5.57) & 1.11(1.11) & 1.10(1.11) & 1.11(1.11) \\
						\midrule
						\multirow{4}[2]{*}{Laplace} & New   & 0.966(0.950) & 0.942(0.940) & 0.950(0.932) & 0.960(0.944) & 0.948(0.950) & 0.944(0.958) \\
						& CYT   & 0.740(0.666) & 0.658(0.658) & 0.666(0.690) & 0.942(0.914) & 0.938(0.936) & 0.914(0.934) \\
						& $\|\hat{\mathbf \Sigma}_1\|_{\infty}$ & 4.10(4.19) & 4.08(4.13) & 4.19(4.11) & 3.58(3.59) & 3.59(3.58) & 3.59(3.60) \\
						& $\|\hat{\mathbf \Sigma}_2\|_{\infty}$ & 5.65(5.66) & 5.66(5.61) & 5.66(5.67) & 1.11(1.12) & 1.12(1.11) & 1.12(1.12) \\
						\bottomrule
					\end{tabular}%
				}
				\label{n=400}%
			\end{table*}%
			
			\begin{figure}[h!] 
				\centering
				\subfigure[Setting 1]{\includegraphics[width=6cm]{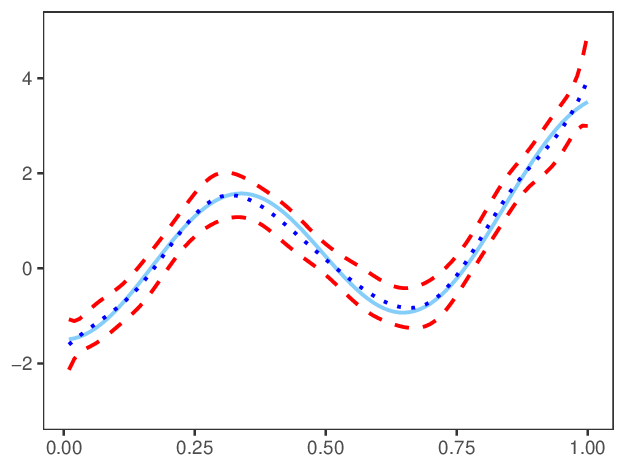}}
				\subfigure[Setting 2]{\includegraphics[width=6cm]{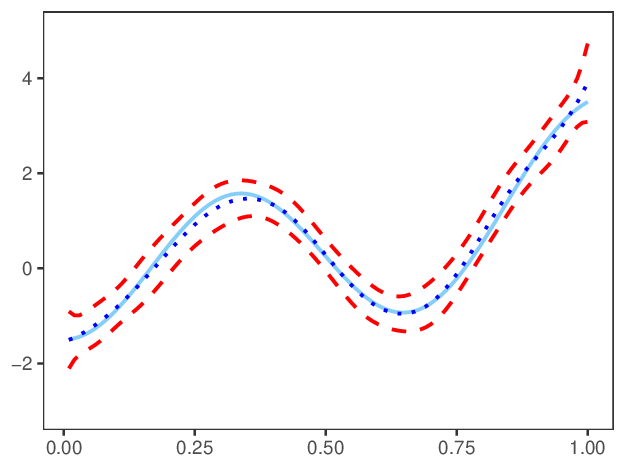}}
				\subfigure[Setting 3]{\includegraphics[width=6cm]{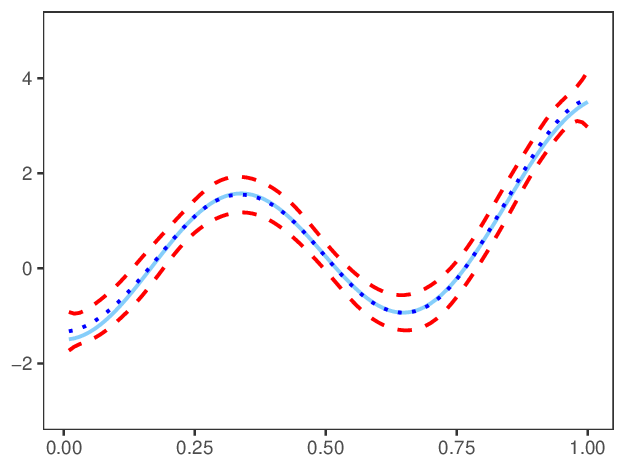}}
				\subfigure[Setting 4]{\includegraphics[width=6cm]{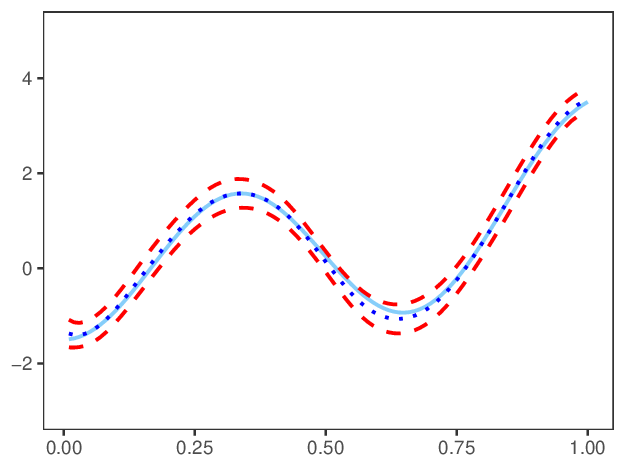}}
				\caption{Numerical simulation with $n=200$. Solid line: the true mean function; dotted line: the estimated mean function using cubic B-splines; dashed line: the proposed SCB at the significance level of $95\%$. }
				\label{n=200figure}
			\end{figure}
			
			\begin{figure}[h!] 
				\centering
				\subfigure[Setting 1]{\includegraphics[width=6cm]{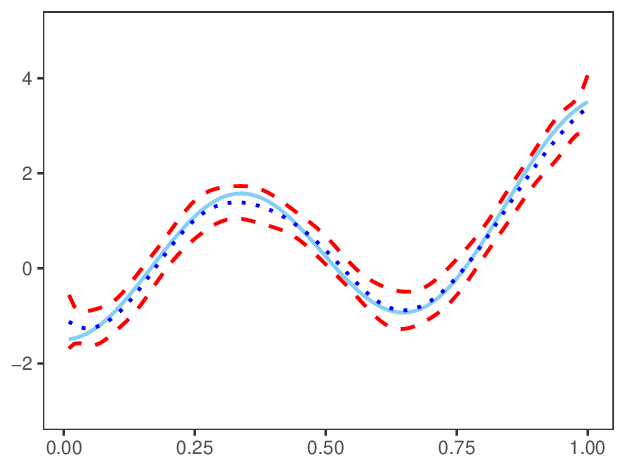}}
				\subfigure[Setting 2]{\includegraphics[width=6cm]{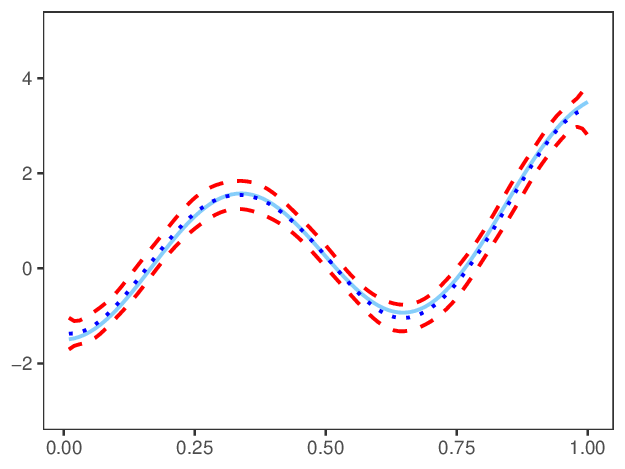}}
				\subfigure[Setting 3]{\includegraphics[width=6cm]{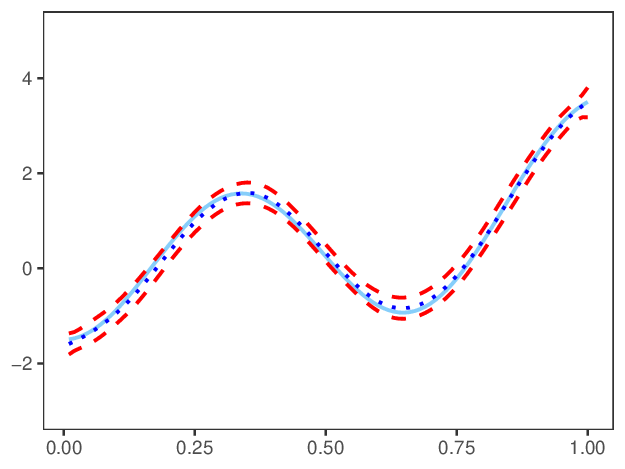}}
				\subfigure[Setting 4]{\includegraphics[width=6cm]{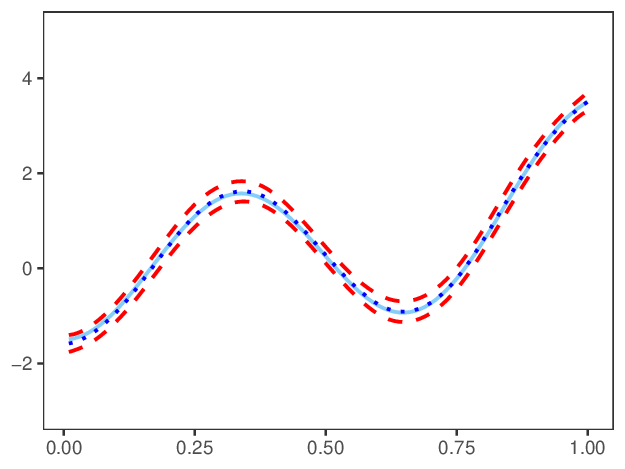}}
				\caption{ Numerical simulation with $n=400$. Solid line: the true mean function; dotted line: the estimated mean function using cubic B-splines; dashed line: the proposed SCB at the significance level of $95\%$. }
				\label{n=400figure}
			\end{figure}
			
			\section{Application}
			We illustrate the proposed method by  empirical analysis of two data-sets.
			\subsection{Sparse case} 
			\par The first data   originated from a prospective study investigating the   accumulation of body fat in a cohort of $162$ girls ($n=162$) participating in the MIT Growth and Development Study, which is available from \url{https://content.sph.harvard.edu/fitzmaur/ala/}; see \cite{phillips2003longitudinal} and \cite{fitzmaurice2012applied} for reference.   The study focused on analyzing changes in percent body fat  before and after menarche. All participants were longitudinally tracked through a schedule of annual measurements until four years  after menarche. During each examination, body fatness was assessed using bioelectric impedance analysis, and the percentage of body fat (\%BF) was derived. The dataset comprises a total of $1049$ individual percent body fat measurements, averaging $6.4$ measurements per subject ($\bar N=6.4$), which is illustrated in Figure \ref{fat-rawdata}.
			\begin{figure}[h!] 
				\centering
				\includegraphics[width=6cm]{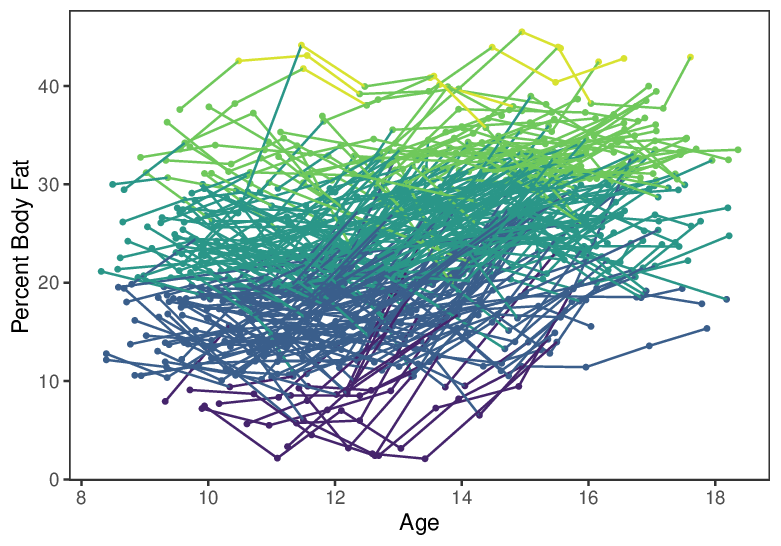}
				\includegraphics[width=6cm]{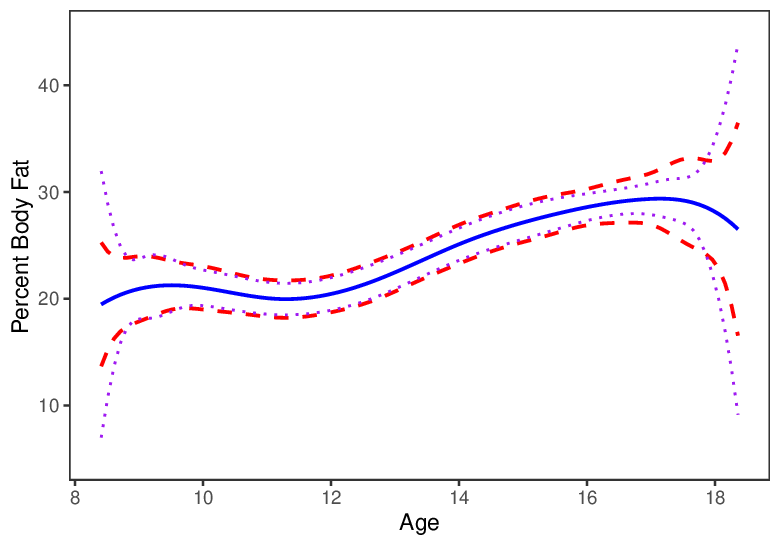}\caption{Left panel: percent body fat of 162 girls before and after menarche. Right panel: solid line: the estimated mean function using cubic B-splines; dashed line: the proposed SCB at the significance level of $95\%$; dotted line: the SCB developed in \cite{cao2012simultaneous} at the significance level of $95\%$.}
				\label{fat-rawdata}
			\end{figure}
			Figure \ref{fat-rawdata} also shows that the estimated mean function of the percent body fat and the corresponding  $95\%$ SCB, which indicates that   percent body fat of girls gradually increases with age between 12 and 16 years old.
			
			\subsection{Dense case}
			The second  dataset is accessible through the University of California, Irvine machine learning repository  \url{https://archive.ics.uci.edu/dataset/157/dodgers+loop+sensor}. It was collected from a loop sensor for the Glendale on ramp for the $101$ North freeway in Los Angeles, located near Dodger Stadium, which
			is the home  of the Los Angeles Dodgers baseball team. The sensor was designed to capture traffic volume, and it recorded measurements of the total number of cars every five minutes.  For  days when the Dodgers had a home game, additional details are available, including the time when the game started and ended.
			
			This traffic data has been studied by \cite{ihler2006adaptive}, \cite{zhang2015varying}, while
			we are interested in   the trend of car counts around the end of games on game days. For this purpose, we consider the end time of each game as
			the zero point ($t=0$) and focus on the car counts between $120$ minutes before the end ($t = -120$) and $120$
			minutes after the game ($t = 120$).
			The data we use  spanned from April 2005 to October 2005 including 78 game days ($n=78$) with $\bar N=48.1$, which is illustrated in Figure \ref{car-rawdata}.
			\begin{figure}[h!] 
				\centering
				\includegraphics[width=6cm]{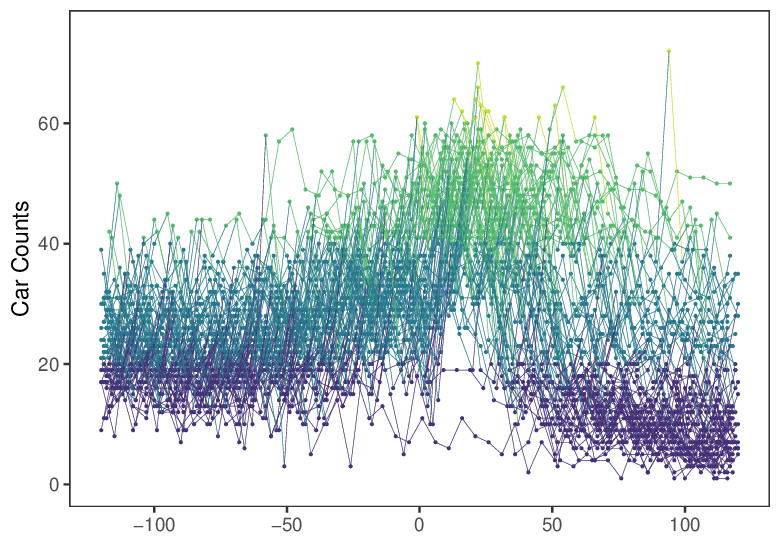}
				\includegraphics[width=6cm]{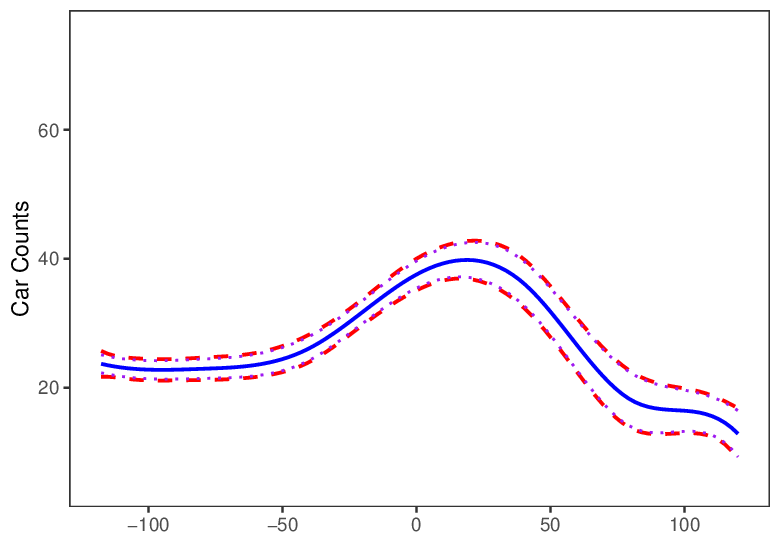}
				\caption{Left panel: car counts on 78 game days over 120 minutes before the end of a game until 120 minutes after
					the game. Right panel: solid line: the estimated mean function using cubic B-splines; dashed line: the proposed SCB at the significance level of $95\%$; dotted line: the SCB developed in \cite{cao2012simultaneous} at the significance level of $95\%$.}
				\label{car-rawdata}
			\end{figure}
			Figure \ref{car-rawdata} also reveals that the car counts exhibits a unimodal shape, steadily rising and reaching its peak approximately $20$ minutes after the game  and subsequently decreasing thereafter.

			\section{Concluding remark}
			\par In this paper, we present a unified framework for the simultaneous inference of mean functions, applicable irrespective of sampling strategy constraints. Our method involves the use of B-splines to construct SCBs, which are shown to be asymptotically valid  from sparse to dense observations. We explore the behavior of the approximated Gaussian process to describe the phase transition phenomenon, balancing between the effect   of  $\Var(Y_{ij})$ and   $\Cov(Y_{ij},Y_{ij^\prime})$. This meticulous analysis leads to categorizing samples as sparse, semi-dense, or dense. Notably, our results align with those in sparse and dense cases as discussed in \cite{zheng2014smooth} and \cite{cao2012simultaneous}, respectively. Moreover, our convergence rate approaches the optimal rate up to a logarithm term, offering a comparison to \cite{cai2012minimax} and \cite{berger2023dense}. 
			
			Furthermore, our methodology extends to orthogonal series estimators, achieving similar results with adjustments based on series-specific parameters. A significant advantage of our proposed SCB lies in its broader applicability compared to other SCBs, which are often constrained by sampling designs. This flexibility is particularly valuable in practical settings where determining the sparseness or denseness of a dataset is not straightforward. 
			
			Additionally, this framework's potential applications extend beyond mean function estimation. It can be adapted for estimating other components, such as covariance functions and eigenfunctions, which involve multi-step nonparametric smoothing. Another promising avenue is its generalization to dependent functional data, which poses the intriguing challenge of estimating long-run covariance functions. These extensions represent both exciting and demanding prospects for future research in this field.

			\section*{Acknowledgements}
			\par This research was supported by the National Natural Science Foundation of China awarded 12171269.
			
			\section*{Supplementary Material}
			\par Supplementary Material contains  detailed proofs of the theoretical results  with necessarily technical lemmas

			\bibliographystyle{apalike} 
			\bibliography{ref}

		\end{document}